\documentclass[10pt]{article}
\usepackage{graphicx}
\usepackage{amsmath}
\usepackage{amssymb}
\usepackage{caption2}
\begin{document}
\begin{center}
{\large {\bf \sc{  Analysis of the hidden-bottom tetraquark mass spectrum with the QCD sum rules }}} \\[2mm]
Zhi-Gang  Wang \footnote{E-mail: zgwang@aliyun.com.  }     \\
 Department of Physics, North China Electric Power University, Baoding 071003, P. R. China
\end{center}

\begin{abstract}
In this article,  we extend our previous work to study the  mass spectrum of the ground state hidden-bottom tetraquark states  with
the QCD sum rules in a systematic way.  The predicted hidden-bottom tetraquark masses can be confronted to the experimental data in the future to diagnose the nature of the $Z_b$ states. In calculations, we observe that the scalar diquark states,  the axialvector diquark states and the axialvector components of the tensor diquark state  are all good diquarks in building the lowest tetraquark states.
 \end{abstract}



\section{Introduction}
In 2011, the Belle collaboration  observed  the $Z^\pm_b(10610)$ and $Z_b^\pm(10650)$ in the $\pi^{\pm}\Upsilon({\rm 1,2,3S})$  and $\pi^{\pm} h_b({\rm 1,2P})$   mass spectrum for the first time, the favored   quantum numbers (Isospin, $G$-parity, Spin, Parity)  are $I^G(J^P)=1^+(1^+)$  \cite{Belle1105}.
Later, the Belle collaboration updated the  values of the masses and widths  $ M_{Z_b(10610)}=(10607.2\pm2.0)\,\rm{ MeV}$, $M_{Z_b(10650)}=(10652.2\pm1.5)\,\rm{MeV}$, $\Gamma_{Z_b(10610)}=(18.4\pm2.4) \,\rm{MeV}$ and
$\Gamma_{Z_b(10650)}=(11.5\pm2.2)\,\rm{ MeV}$ \cite{Belle1110}, which are adopted in the  Review of Particle Physics by the Particle Data Group now \cite{PDG}.
The possible assignments of the $Z^\pm_b(10610)$ and $Z_b^\pm(10650)$ are the  tetraquark states \cite{Tetraquark-Zb,Tetraquark-Zb-QCDSR,WangHuang-2014-NPA}, molecular states \cite{Molecule-Zb-A,Molecule-Zb-B,Molecule-Zb-C,Molecule-Zb-D,Molecule-Zb-E}, threshold cusps \cite{Cusp-Zb}, re-scattering effects \cite{Rescatter-Zb}, etc.
For more literatures  on the $Z_b$ states, one can consult the old review \cite{FKGuo-EPJA} and the recent review \cite{LiuYR-1903}.

In 2013, the BESIII collaboration (also the Belle collaboration) observed the $Z_c^{\pm}(3900)$ in the $\pi^\pm J/\psi$ mass spectrum \cite{BES3900,Belle3900}.
   Later, the BESIII collaboration  observed
the $Z^{\pm}_c(4025)$ near the $(D^{*} \bar{D}^{*})^{\pm}$ threshold \cite{BES1308},
and the  $Z_c^{\pm}(4020)$   in the $\pi^\pm h_c$ mass spectrum  \cite{BES1309}. Now the $Z_c^{\pm}(4025)$ and
$Z_c^{\pm}(4020)$ are taken to be the same particle, and are denoted as the $X(4020)$ in the Review of Particle Physics \cite{PDG}.

The $Z_c^\pm(3900)$ and $Z_c^\pm(4020)$ ($Z^\pm_b(10610)$ and $Z_b^\pm(10650)$) are charged charmonium-like states (bottomonium-like states), their quark constituents must be $c\bar{c}u\bar{d}$ or $c\bar{c}d\bar{u}$ ($b\bar{b}u\bar{d}$ or $b\bar{b}d\bar{u}$), irrespective of the diquark-antidiquark type or meson-meson type substructures.
The $Z_b(10610)$, $Z_b(10650)$, $Z_c(3900)$ and $Z_c(4020)$ are observed in the analogous decays to the final states  $\pi^{\pm} h_b({\rm 1,2P})$, $\pi^{\pm}\Upsilon({\rm 1,2,3S})$, $\pi^\pm J/\psi$, $\pi^\pm h_c$, and should have analogous structures.
In Refs.\cite{WangHuang-2014-NPA,WangHuangtao-2014-PRD,Wang-tetra-formula}, we assign the $Z_b(10610)$, $Z_b(10650)$, $Z_c(3900)$ and $Z_c(4020)$
to be the diquark-antidiquark type  axialvector tetraquark states, and study their masses   with the QCD sum rules in details.  Furthermore, we  explore the energy scale dependence of the hidden-charm and hidden-bottom tetraquark states   for the first time \cite{WangHuangtao-2014-PRD}, and suggest a  formula,
\begin{eqnarray}
\mu&=&\sqrt{M^2_{X/Y/Z}-(2{\mathbb{M}}_Q)^2} \, ,
 \end{eqnarray}
 with the effective heavy  mass ${\mathbb{M}}_Q$ to determine the optimal energy scales $\mu$ of the  QCD spectral densities \cite{WangHuang-2014-NPA,Wang-tetra-formula}. The experimental values of the masses can be well reproduced. In Ref.\cite{WangZhang-Solid}, we  study the two-body strong decays $Z_c^\pm(3900)\to J/\psi\pi^\pm$, $\eta_c\rho^\pm$, $(D \bar{D}^{*})^\pm$  with the QCD sum rules in details. We take into account both the connected and disconnected Feynman diagrams, and pay special attentions  to matching the hadron side  with the QCD side of the correlation functions to obtain solid duality, the predicted width $\Gamma_{Z_c(3900)}=54.2\pm29.8\,\rm{MeV}$ is consistent with the experimental data \cite{BES3900,Belle3900},  and supports assigning the $Z_c^\pm(3900)$ to be  the diquark-antidiquark  type axialvector  tetraquark  state \cite{WangHuangtao-2014-PRD}. In Ref.\cite{WangHuang-2014-NPA}, we use  the method proposed in Ref.\cite{Nielsen3900} to study the two-body strong decays $Z_b^\pm(10610)\to \Upsilon\pi^\pm$, $\Upsilon\rho^\pm$ with the QCD sum rules by taking into account only the connected Feynman diagrams. Although the predictions are
 good,  the subtractions of the  higher resonances  and continuum states are introduced by hand, the contaminations  cannot be subtracted  completely. The widths of the $Z_b^\pm(10610)$ and $Z_b^\pm(10650)$ should be studied in a consistent way to make the assignments more robust. A updated analysis of the masses and widths with the QCD sum rules is needed.

If the $Z_b^\pm(10610)$ and $Z_b^\pm(10650)$ are the diquark-antidiquark type axialvector hidden-bottom tetraquark states, there should exist a holonomic spectrum for the scalar, axialvector and tensor hidden-bottom tetraquark states without introducing an additional P-wave. Now we extend our previous work to study the mass spectrum of the hidden-bottom tetraquark states in a systematic way. Those hidden-bottom tetraquark states may be observed  at the LHCb, Belle II,  CEPC (Circular Electron Positron Collider), FCC (Future Circular Collider), ILC (International Linear Collider)  in the future, and shed light on the nature of the exotic $X$, $Y$, $Z$ particles.

We usually take the diquarks (in color antitriplet) and antidiquarks (in color triplet) as the basic building blocks to construct  the tetraquark states.   The diquarks (or diquark operators) $\varepsilon^{abc}q^{T}_b C\Gamma q^{\prime}_c$  have  five  structures  in Dirac spinor space, where $C\Gamma=C\gamma_5$, $C$, $C\gamma_\mu \gamma_5$,  $C\gamma_\mu $ and $C\sigma_{\mu\nu}$ (or $C\sigma_{\mu\nu}\gamma_5$) for the scalar ($S$), pseudoscalar ($P$), vector ($V$), axialvector ($A$)  and  tensor ($T$) diquarks, respectively, the $a$, $b$, $c$ are color indexes. The scalar, pseudoscalar, vector and axialvector  diquark states have  been studied with the QCD sum rules in details,   the good diquark correlations in building the lowest tetraquark states   are the scalar  and axialvector  diquark states \cite{Dosch-Diquark-1989,WangDiquark}, the axialvector  diquark states are not bad diquark states.

 Under parity transform $\widehat{P}$, the tensor diquark  operators have the  properties,
\begin{eqnarray}
\widehat{P}\varepsilon^{abc}q^{Tb}(x)C\sigma_{\mu\nu}\gamma_5Q^c(x)\widehat{P}^{-1}&=&\varepsilon^{abc}q^{Tb}(\tilde{x})C\sigma^{\mu\nu}\gamma_5Q^c(\tilde{x}) \, , \nonumber\\
\widehat{P}\varepsilon^{abc}q^{Tb}(x)C\sigma_{\mu\nu} Q^c(x)\widehat{P}^{-1}&=&-\varepsilon^{abc}q^{Tb}(\tilde{x})C\sigma^{\mu\nu} Q^c(\tilde{x}) \, ,
\end{eqnarray}
where  $x^\mu=(t,\vec{x})$ and $\tilde{x}^\mu=(t,-\vec{x})$. The tensor diquark states have both $J^P=1^+$ and $1^-$ components, we introduce the four vector $t^\mu=(1,\vec{0})$ and project out the $1^+$ and $1^-$ components explicitly,
\begin{eqnarray}
\widehat{P}\varepsilon^{abc}q^{Tb}(x)C\sigma^t_{\mu\nu}\gamma_5Q^c(x)\widehat{P}^{-1}&=&+\varepsilon^{abc}q^{Tb}(\tilde{x})C\sigma^t_{\mu\nu}\gamma_5Q^c(\tilde{x}) \, , \nonumber\\
\widehat{P}\varepsilon^{abc}q^{Tb}(x)C\sigma^v_{\mu\nu} Q^c(x)\widehat{P}^{-1}&=&+\varepsilon^{abc}q^{Tb}(\tilde{x})C\sigma^v_{\mu\nu} Q^c(\tilde{x}) \, , \nonumber\\
\widehat{P}\varepsilon^{abc}q^{Tb}(x)C\sigma^t_{\mu\nu} Q^c(x)\widehat{P}^{-1}&=&-\varepsilon^{abc}q^{Tb}(\tilde{x})C\sigma^t_{\mu\nu} Q^c(\tilde{x}) \, , \nonumber\\
\widehat{P}\varepsilon^{abc}q^{Tb}(x)C\sigma^v_{\mu\nu}\gamma_5Q^c(x)\widehat{P}^{-1}&=&-\varepsilon^{abc}q^{Tb}(\tilde{x})C\sigma^v_{\mu\nu}\gamma_5Q^c(\tilde{x}) \, ,
\end{eqnarray}
where
$\sigma^t_{\mu\nu} =\frac{i}{2}\Big[\gamma^t_\mu, \gamma^t_\nu \Big]$, $\sigma^v_{\mu\nu} =\frac{i}{2}\Big[\gamma^v_\mu, \gamma^t_\nu \Big]$,
$\gamma^v_\mu =  \gamma \cdot t t_\mu$, $\gamma^t_\mu=\gamma_\mu-\gamma \cdot t t_\mu$ \cite{WangZG-Z4100-0903}. 
Thereafter,  we will denote the axialvector diquark operators $\varepsilon^{abc}q^{Tb}(x)C\sigma^v_{\mu\nu} Q^c(x)$, $\varepsilon^{abc}q^{Tb}(x)C\sigma^t_{\mu\nu} \gamma_5Q^c(x)$ as $\widetilde{A}$,  and the vector diquark operators $\varepsilon^{abc}q^{Tb}(x)C\sigma^t_{\mu\nu} Q^c(x)$,  $\varepsilon^{abc}q^{Tb}(x)C\sigma^v_{\mu\nu} \gamma_5Q^c(x)$ as $\widetilde{V}$.

In this article, we take the scalar ($S$), axialvector ($A$, $\widetilde{A}$), vector ($V$, $\widetilde{V}$)  diquark operators and antidiquark operators as the basic building blocks    to construct the tetraquark operators  to study the mass spectrum of the hidden-bottom tetraquark states with the QCD sum rules in a systematic way.

The article is arranged as follows:  we derive the QCD sum rules for the masses and pole residues  of  the  hidden-bottom tetraquark states in section 2; in section 3, we   present the numerical results and discussions; section 4 is reserved for our conclusion.

\section{QCD sum rules for  the  hidden-bottom tetraquark states}

We write down  the two-point correlation functions $\Pi(p)$, $\Pi_{\mu\nu}(p)$ and $\Pi_{\mu\nu\alpha\beta}(p)$  in the QCD sum rules,
\begin{eqnarray}
\Pi(p)&=&i\int d^4x e^{ip \cdot x} \langle0|T\Big\{J(x)J^{\dagger}(0)\Big\}|0\rangle \, ,\nonumber\\
\Pi_{\mu\nu}(p)&=&i\int d^4x e^{ip \cdot x} \langle0|T\Big\{J_\mu(x)J_{\nu}^{\dagger}(0)\Big\}|0\rangle \, ,\nonumber\\
\Pi_{\mu\nu\alpha\beta}(p)&=&i\int d^4x e^{ip \cdot x} \langle0|T\Big\{J_{\mu\nu}(x)J_{\alpha\beta}^{\dagger}(0)\Big\}|0\rangle \, ,
\end{eqnarray}
where the currents $J(x)=J_{SS}(x)$,  $J_{AA}(x)$, $J_{\widetilde{A}\widetilde{A}}(x)$,
$J_{\widetilde{V}\widetilde{V}}(x)$, $J_\mu(x)=J^{SA}_{-,\mu}(x)$, $J_{-,\mu}^{\widetilde{A}A}(x)$,  $J^{SA}_{+,\mu}(x)$,
$J_{+,\mu}^{\widetilde{V}V}(x)$, $J_{+,\mu}^{\widetilde{A}A}(x)$,
$J_{\mu\nu}(x)=J^{AA}_{-,\mu\nu}(x)$, $J^{S\widetilde{A}}_{-,\mu\nu}(x)$, $J^{AA}_{+,\mu\nu}(x)$,
\begin{eqnarray}
J_{SS}(x)&=&\varepsilon^{ijk}\varepsilon^{imn}u^{Tj}(x)C\gamma_5 b^k(x)  \bar{d}^m(x)\gamma_5 C \bar{b}^{Tn}(x) \, ,\nonumber \\
J_{AA}(x)&=&\varepsilon^{ijk}\varepsilon^{imn}u^{Tj}(x)C\gamma_\mu b^k(x)  \bar{d}^m(x)\gamma^\mu C \bar{b}^{Tn}(x) \, ,\nonumber \\
J_{\tilde{A}\tilde{A}}(x)&=&\varepsilon^{ijk}\varepsilon^{imn}u^{Tj}(x)C\sigma^v_{\mu\nu} b^k(x)  \bar{d}^m(x)\sigma_v^{\mu\nu} C \bar{b}^{Tn}(x) \, ,\nonumber \\
J_{\tilde{V}\tilde{V}}(x)&=&\varepsilon^{ijk}\varepsilon^{imn}u^{Tj}(x)C\sigma^t_{\mu\nu} b^k(x)  \bar{d}^m(x)\sigma_t^{\mu\nu} C \bar{b}^{Tn}(x) \, ,\nonumber \\
J^{SA}_{-,\mu}(x)&=&\frac{\varepsilon^{ijk}\varepsilon^{imn}}{\sqrt{2}}\Big[u^{Tj}(x)C\gamma_5b^k(x) \bar{d}^m(x)\gamma_\mu C \bar{b}^{Tn}(x)-u^{Tj}(x)C\gamma_\mu b^k(x)\bar{d}^m(x)\gamma_5C \bar{b}^{Tn}(x) \Big] \, ,\nonumber\\
J^{AA}_{-,\mu\nu}(x)&=&\frac{\varepsilon^{ijk}\varepsilon^{imn}}{\sqrt{2}}\Big[u^{Tj}(x) C\gamma_\mu b^k(x) \bar{d}^m(x) \gamma_\nu C \bar{b}^{Tn}(x)  -u^{Tj}(x) C\gamma_\nu b^k(x) \bar{d}^m(x) \gamma_\mu C \bar{b}^{Tn}(x) \Big] \, , \nonumber\\
J_{-,\mu}^{\widetilde{A}A}(x)&=&\frac{\varepsilon^{ijk}\varepsilon^{imn}}{\sqrt{2}}\Big[u^{Tj}(x)C\sigma_{\mu\nu}\gamma_5 b^k(x)\bar{d}^m(x)\gamma^\nu C \bar{b}^{Tn}(x)-u^{Tj}(x)C\gamma^\nu b^k(x)\bar{d}^m(x)\gamma_5\sigma_{\mu\nu} C \bar{b}^{Tn}(x) \Big] \, , \nonumber\\
J^{S\widetilde{A}}_{-,\mu\nu}(x)&=&\frac{\varepsilon^{ijk}\varepsilon^{imn}}{\sqrt{2}}\Big[u^{Tj}(x)C\gamma_5 b^k(x)  \bar{d}^m(x)\sigma_{\mu\nu} C \bar{b}^{Tn}(x)- u^{Tj}(x)C\sigma_{\mu\nu} b^k(x)  \bar{d}^m(x)\gamma_5 C \bar{b}^{Tn}(x) \Big] \, , \nonumber\\
J^{SA}_{+,\mu}(x)&=&\frac{\varepsilon^{ijk}\varepsilon^{imn}}{\sqrt{2}}\Big[u^{Tj}(x)C\gamma_5b^k(x) \bar{d}^m(x)\gamma_\mu C \bar{b}^{Tn}(x)+u^{Tj}(x)C\gamma_\mu b^k(x)\bar{d}^m(x)\gamma_5C \bar{b}^{Tn}(x) \Big] \, ,\nonumber\\
J_{+,\mu}^{\widetilde{V}V}(x)&=&\frac{\varepsilon^{ijk}\varepsilon^{imn}}{\sqrt{2}}\left[u^{Tj}(x)C\sigma_{\mu\nu} b^k(x)\bar{d}^m(x)\gamma_5\gamma^\nu C \bar{b}^{Tn}(x)-u^{Tj}(x)C\gamma^\nu \gamma_5b^k(x)\bar{d}^m(x) \sigma_{\mu\nu} C \bar{b}^{Tn}(x) \right] \, , \nonumber\\
J_{+,\mu}^{\widetilde{A}A}(x)&=&\frac{\varepsilon^{ijk}\varepsilon^{imn}}{\sqrt{2}}\left[u^{Tj}(x)C\sigma_{\mu\nu}\gamma_5 b^k(x)\bar{d}^m(x)\gamma^\nu C \bar{b}^{Tn}(x)+u^{Tj}(x)C\gamma^\nu b^k(x)\bar{d}^m(x)\gamma_5\sigma_{\mu\nu} C \bar{b}^{Tn}(x) \right] \, , \nonumber\\
J^{AA}_{+,\mu\nu}(x)&=&\frac{\varepsilon^{ijk}\varepsilon^{imn}}{\sqrt{2}}\Big[u^{Tj}(x) C\gamma_\mu b^k(x) \bar{d}^m(x) \gamma_\nu C \bar{b}^{Tn}(x)  +u^{Tj}(x) C\gamma_\nu b^k(x) \bar{d}^m(x) \gamma_\mu C \bar{b}^{Tn}(x) \Big] \, ,
\end{eqnarray}
  the $i$, $j$, $k$, $m$, $n$ are  color indexes, the $C$ is the charge conjugation matrix, the subscripts $\pm$ denote the positive charge  conjugation and negative charge conjugation, respectively. The diquark operators
  $\varepsilon^{ijk}q^{Tj}(x)C\sigma_{\mu\nu}\gamma_5b^j(x)$ and $ \varepsilon^{ijk}q^{Tj}(x)C\sigma_{\mu\nu} b^k(x)$
  have both positive parity and negative parity components, we project out the $J^{P}=1^+$ and $1^-$ components unambiguously with suitable  diquark operators to obtain the current operators  $J_{-,\mu}^{\widetilde{A}A}(x)$,  $J_{+,\mu}^{\widetilde{V}V}(x)$ and $J_{+,\mu}^{\widetilde{A}A}(x)$ with the $J^P=1^+$. The current operators  $J^{AA}_{-,\mu\nu}(x)$ and $J^{S\widetilde{A}}_{-,\mu\nu}(x)$ couple potentially  to both the positive parity and negative parity tetraquark states, we separate those contributions explicitly to obtain reliable QCD sum rules.    In Table \ref{Current-Table}, we present the quark structures and corresponding interpolating currents for the hidden-bottom tetraquark states.  The four vector $t^\mu=(1,\vec{0})$ breaks down Lorentz
covariance, the currents $J_{\widetilde{A}\widetilde{A}}(x)$, $J_{\widetilde{V}\widetilde{V}}(x)$, 
$J_{-,\mu}^{\widetilde{A}A}(x)$, $J^{S\widetilde{A}}_{-,\mu\nu}(x)$, $J_{+,\mu}^{\widetilde{V}V}(x)$ and  
$J_{+,\mu}^{\widetilde{A}A}(x)$ are not Lorentz covariant, it is the shortcoming of the present method, the calculations can be understood as carried out at a particular (or given)  coordinate system, which cannot impair the predictive ability.

\begin{table}
\begin{center}
\begin{tabular}{|c|c|c|c|c|c|c|c|c|}\hline\hline
 $Z_b$                                                                            & $J^{PC}$  & Currents              \\ \hline

$[ub]_{S}[\overline{db}]_{S}$                                                     & $0^{++}$  & $J_{SS}(x)$              \\

$[ub]_{A}[\overline{db}]_{A}$                                                     & $0^{++}$  & $J_{AA}(x)$               \\

$[ub]_{\tilde{A}}[\overline{db}]_{\tilde{A}}$                                     & $0^{++}$  & $J_{\widetilde{A}\widetilde{A}}(x)$             \\

$[ub]_{\tilde{V}}[\overline{db}]_{\tilde{V}}$                                     & $0^{++}$  & $J_{\widetilde{V}\widetilde{V}}(x)$           \\

$[ub]_S[\overline{db}]_{A}-[ub]_{A}[\overline{db}]_S$                             & $1^{+-}$  & $J^{SA}_{-,\mu}(x)$         \\

$[ub]_{A}[\overline{db}]_{A}$                                                     & $1^{+-}$  & $J^{AA}_{-,\mu\nu}(x)$        \\

$[ub]_{\widetilde{A}}[\overline{db}]_{A}-[ub]_{A}[\overline{db}]_{\widetilde{A}}$ & $1^{+-}$  & $J_{-,\mu}^{\widetilde{A}A}(x)$   \\

$[ub]_S[\overline{db}]_{\widetilde{A}}-[ub]_{\widetilde{A}}[\overline{db}]_S$     & $1^{+-}$  & $J^{S\widetilde{A}}_{-,\mu\nu}(x)$     \\

$[ub]_S[\overline{db}]_{A}+[ub]_{A}[\overline{db}]_S$                             & $1^{++}$  & $J^{SA}_{+,\mu}(x)$        \\

$[ub]_{\widetilde{V}}[\overline{db}]_{V}-[ub]_{V}[\overline{db}]_{\widetilde{V}}$ & $1^{++}$  & $J_{+,\mu}^{\widetilde{V}V}(x)$      \\

$[ub]_{\widetilde{A}}[\overline{db}]_{A}+[ub]_{A}[\overline{db}]_{\widetilde{A}}$ & $1^{++}$  & $J_{+,\mu}^{\widetilde{A}A}(x)$       \\

$[ub]_{A}[\overline{db}]_{A}$                                                     & $2^{++}$  & $J^{AA}_{+,\mu\nu}(x)$       \\ \hline\hline
\end{tabular}
\end{center}
\caption{ The quark structures and corresponding current operators  for the hidden-bottom tetraquark states. }\label{Current-Table}
\end{table}

At the hadron side, we  insert  a complete set of intermediate hadronic states with
the same quantum numbers as the current operators $J(x)$, $J_\mu(x)$ and $J_{\mu\nu}(x)$ into the
correlation functions $\Pi(p)$, $\Pi_{\mu\nu}(p)$ and $\Pi_{\mu\nu\alpha\beta}(p)$   to obtain the hadronic representation
\cite{SVZ79,Reinders85}, and isolate the ground state hidden-bottom tetraquark contributions,
\begin{eqnarray}
\Pi(p)&=&\frac{\lambda_{Z^+}^2}{M_{Z^+}^2-p^2} +\cdots \nonumber\\
&=&\Pi_{+}(p^2) \, ,\nonumber\\
\Pi_{\mu\nu}(p)&=&\frac{\lambda_{Z^+}^2}{M_{Z^+}^2-p^2}\left( -g_{\mu\nu}+\frac{p_{\mu}p_{\nu}}{p^2}\right) +\cdots \nonumber\\
&=&\Pi_{+}(p^2)\left( -g_{\mu\nu}+\frac{p_{\mu}p_{\nu}}{p^2}\right)+\cdots \, ,\nonumber\\
\Pi^{AA,-}_{\mu\nu\alpha\beta}(p)&=&\frac{\lambda_{ Z^+}^2}{M_{Z^+}^2\left(M_{Z^+}^2-p^2\right)}\left(p^2g_{\mu\alpha}g_{\nu\beta} -p^2g_{\mu\beta}g_{\nu\alpha} -g_{\mu\alpha}p_{\nu}p_{\beta}-g_{\nu\beta}p_{\mu}p_{\alpha}+g_{\mu\beta}p_{\nu}p_{\alpha}+g_{\nu\alpha}p_{\mu}p_{\beta}\right) \nonumber\\
&&+\frac{\lambda_{ Z^-}^2}{M_{Z^-}^2\left(M_{Z^-}^2-p^2\right)}\left( -g_{\mu\alpha}p_{\nu}p_{\beta}-g_{\nu\beta}p_{\mu}p_{\alpha}+g_{\mu\beta}p_{\nu}p_{\alpha}+g_{\nu\alpha}p_{\mu}p_{\beta}\right) +\cdots  \nonumber\\
&=&\widetilde{\Pi}_{+}(p^2)\left(p^2g_{\mu\alpha}g_{\nu\beta} -p^2g_{\mu\beta}g_{\nu\alpha} -g_{\mu\alpha}p_{\nu}p_{\beta}-g_{\nu\beta}p_{\mu}p_{\alpha}+g_{\mu\beta}p_{\nu}p_{\alpha}+g_{\nu\alpha}p_{\mu}p_{\beta}\right) \nonumber\\
&&+\widetilde{\Pi}_{-}(p^2)\left( -g_{\mu\alpha}p_{\nu}p_{\beta}-g_{\nu\beta}p_{\mu}p_{\alpha}+g_{\mu\beta}p_{\nu}p_{\alpha}+g_{\nu\alpha}p_{\mu}p_{\beta}\right) \, ,\nonumber\\
\Pi^{S\widetilde{A},-}_{\mu\nu\alpha\beta}(p)&=&\frac{\lambda_{ Z^-}^2}{M_{Z^-}^2\left(M_{Z^-}^2-p^2\right)}\left(p^2g_{\mu\alpha}g_{\nu\beta} -p^2g_{\mu\beta}g_{\nu\alpha} -g_{\mu\alpha}p_{\nu}p_{\beta}-g_{\nu\beta}p_{\mu}p_{\alpha}+g_{\mu\beta}p_{\nu}p_{\alpha}+g_{\nu\alpha}p_{\mu}p_{\beta}\right) \nonumber\\
&&+\frac{\lambda_{ Z^+}^2}{M_{Z^+}^2\left(M_{Z^+}^2-p^2\right)}\left( -g_{\mu\alpha}p_{\nu}p_{\beta}-g_{\nu\beta}p_{\mu}p_{\alpha}+g_{\mu\beta}p_{\nu}p_{\alpha}+g_{\nu\alpha}p_{\mu}p_{\beta}\right) +\cdots  \nonumber\\
&=&\widetilde{\Pi}_{-}(p^2)\left(p^2g_{\mu\alpha}g_{\nu\beta} -p^2g_{\mu\beta}g_{\nu\alpha} -g_{\mu\alpha}p_{\nu}p_{\beta}-g_{\nu\beta}p_{\mu}p_{\alpha}+g_{\mu\beta}p_{\nu}p_{\alpha}+g_{\nu\alpha}p_{\mu}p_{\beta}\right) \nonumber\\
&&+\widetilde{\Pi}_{+}(p^2)\left( -g_{\mu\alpha}p_{\nu}p_{\beta}-g_{\nu\beta}p_{\mu}p_{\alpha}+g_{\mu\beta}p_{\nu}p_{\alpha}+g_{\nu\alpha}p_{\mu}p_{\beta}\right) \, , \nonumber\\
\Pi_{\mu\nu\alpha\beta}^{AA,+}(p)&=&\frac{\lambda_{ Z^+}^2}{M_{Z^+}^2-p^2}\left( \frac{\widetilde{g}_{\mu\alpha}\widetilde{g}_{\nu\beta}+\widetilde{g}_{\mu\beta}\widetilde{g}_{\nu\alpha}}{2}-\frac{\widetilde{g}_{\mu\nu}\widetilde{g}_{\alpha\beta}}{3}\right) +\cdots \, \, , \nonumber \\
&=&\Pi_{+}(p^2)\left( \frac{\widetilde{g}_{\mu\alpha}\widetilde{g}_{\nu\beta}+\widetilde{g}_{\mu\beta}\widetilde{g}_{\nu\alpha}}{2}-\frac{\widetilde{g}_{\mu\nu}\widetilde{g}_{\alpha\beta}}{3}\right) +\cdots\, ,
\end{eqnarray}
where $\widetilde{g}_{\mu\nu}=g_{\mu\nu}-\frac{p_{\mu}p_{\nu}}{p^2}$, the superscripts (subscripts) $\pm$ in the hidden-bottom tetraquark states $Z_b^{\pm}$ (components $\Pi_{\pm}(p^2)$, $\widetilde{\Pi}_{\pm}(p^2)$) denote the positive-parity and negative parity, respectively. The pole residues $\lambda_{Z^\pm}$ are defined by
\begin{eqnarray}
 \langle 0|J(0)|Z_b^+(p)\rangle &=&\lambda_{Z^+}\, , \nonumber\\
 \langle 0|J_\mu(0)|Z_b^+(p)\rangle &=&\lambda_{Z^+}\varepsilon_\mu\, , \nonumber\\
  \langle 0|J_{-,\mu\nu}^{S\widetilde{A}}(0)|Z_b^-(p)\rangle &=& \frac{\lambda_{Z^-}}{M_{Z^-}} \, \varepsilon_{\mu\nu\alpha\beta} \, \varepsilon^{\alpha}p^{\beta}\, , \nonumber\\
 \langle 0|J_{-,\mu\nu}^{S\widetilde{A}}(0)|Z_b^+(p)\rangle &=&\frac{\lambda_{Z^+}}{M_{Z^+}} \left(\varepsilon_{\mu}p_{\nu}-\varepsilon_{\nu}p_{\mu} \right)\, , \nonumber\\
  \langle 0|J_{-,\mu\nu}^{AA}(0)|Z_b^+(p)\rangle &=& \frac{\lambda_{Z^+}}{M_{Z^+}} \, \varepsilon_{\mu\nu\alpha\beta} \, \varepsilon^{\alpha}p^{\beta}\, , \nonumber\\
 \langle 0|J_{-,\mu\nu}^{AA}(0)|Z_b^-(p)\rangle &=&\frac{\lambda_{Z^-}}{M_{Z^-}} \left(\varepsilon_{\mu}p_{\nu}-\varepsilon_{\nu}p_{\mu} \right)\, , \nonumber\\
  \langle 0|J_{+,\mu\nu}^{AA}(0)|Z_b^+(p)\rangle &=& \lambda_{Z^+}\, \varepsilon_{\mu\nu} \, ,
\end{eqnarray}
the  $\varepsilon_{\mu/\alpha}$ and $\varepsilon_{\mu\nu}$ are the polarization vectors of the hidden-bottom tetraquark states.
In this article, we choose the components $\Pi_{+}(p^2)$ and $p^2\widetilde{\Pi}_{+}(p^2)$ to study the scalar, axialvector and tensor hidden-bottom tetraquark states with the QCD sum rules.

At the QCD side, we carry out the operator product expansion for the correlation functions $\Pi(p)$, $\Pi_{\mu\nu}(p)$ and $\Pi_{\mu\nu\alpha\beta}(p)$ up to the vacuum condensates of dimension $10$ in a consistent way, and obtain the QCD spectral densities $\rho(s)$ through dispersion relation. We match  the hadron side with the QCD  side of the correlation functions  below the continuum threshold  $s_0$ and perform Borel transform  with respect to
 $P^2=-p^2$ to obtain  the  QCD sum rules:
\begin{eqnarray}\label{QCDSR}
\lambda^2_{Z^+}\, \exp\left(-\frac{M^2_{Z^+}}{T^2}\right)= \int_{4m_b^2}^{s_0} ds\, \rho(s) \, \exp\left(-\frac{s}{T^2}\right) \, .
\end{eqnarray}
 The explicit expressions of the QCD spectral densities $\rho(s)$ are available upon request by contacting me via  E-mail. For the technical details, one can consult Refs.\cite{WangHuangtao-2014-PRD,Wang-tetra-formula}.

We derive Eq.\eqref{QCDSR} with respect to  $\tau=\frac{1}{T^2}$,  and obtain the QCD sum rules for
 the masses of the  scalar,  axialvector  and tensor hidden-bottom tetraquark states $Z_b$ through a ratio,
 \begin{eqnarray}
 M^2_{Z^+}&=& -\frac{\int_{4m_b^2}^{s_0} ds\frac{d}{d \tau}\rho(s)\exp\left(-\tau s \right)}{\int_{4m_b^2}^{s_0} ds \rho(s)\exp\left(-\tau s\right)}\, .
\end{eqnarray}

\section{Numerical results and discussions}
We take  the standard values of the vacuum condensates $\langle
\bar{q}q \rangle=-(0.24\pm 0.01\, \rm{GeV})^3$,   $\langle
\bar{q}g_s\sigma G q \rangle=m_0^2\langle \bar{q}q \rangle$,
$m_0^2=(0.8 \pm 0.1)\,\rm{GeV}^2$,  $\langle \frac{\alpha_s
GG}{\pi}\rangle=(0.33\,\rm{GeV})^4 $    at the energy scale  $\mu=1\, \rm{GeV}$
\cite{SVZ79,Reinders85,Colangelo-Review}, and take the $\overline{MS}$ mass $m_{b}(m_b)=(4.18\pm0.03)\,\rm{GeV}$ from the Particle Data Group \cite{PDG}, and set $m_u=m_d=0$.
Furthermore, we take into account the energy-scale dependence of  the input parameters at  the QCD side,
\begin{eqnarray}
\langle\bar{q}q \rangle(\mu)&=&\langle\bar{q}q \rangle({\rm 1GeV})\left[\frac{\alpha_{s}({\rm 1GeV})}{\alpha_{s}(\mu)}\right]^{\frac{12}{33-2n_f}}\, , \nonumber\\
 \langle\bar{q}g_s \sigma Gq \rangle(\mu)&=&\langle\bar{q}g_s \sigma Gq \rangle({\rm 1GeV})\left[\frac{\alpha_{s}({\rm 1GeV})}{\alpha_{s}(\mu)}\right]^{\frac{2}{33-2n_f}}\, , \nonumber\\
 m_b(\mu)&=&m_b(m_b)\left[\frac{\alpha_{s}(\mu)}{\alpha_{s}(m_b)}\right]^{\frac{12}{33-2n_f}} \, ,\nonumber\\
\alpha_s(\mu)&=&\frac{1}{b_0t}\left[1-\frac{b_1}{b_0^2}\frac{\log t}{t} +\frac{b_1^2(\log^2{t}-\log{t}-1)+b_0b_2}{b_0^4t^2}\right]\, ,
\end{eqnarray}
   where $t=\log \frac{\mu^2}{\Lambda^2}$, $b_0=\frac{33-2n_f}{12\pi}$, $b_1=\frac{153-19n_f}{24\pi^2}$, $b_2=\frac{2857-\frac{5033}{9}n_f+\frac{325}{27}n_f^2}{128\pi^3}$,  $\Lambda=210\,\rm{MeV}$, $292\,\rm{MeV}$  and  $332\,\rm{MeV}$ for the flavors  $n_f=5$, $4$ and $3$, respectively  \cite{PDG,Narison-mix}, and evolve all the input parameters to the optimal  energy scales   $\mu$ with the flavor $n_f=5$ to extract the tetraquark masses.

In all the QCD sum rules for the hidden-charm tetraquark states \cite{Wang-tetra-formula,Wang-tetra-Zc4430}, hidden-charm pentaquark states \cite{WangPentaQuark} and hidden-bottom tetraquark states \cite{WangHuang-2014-NPA}, we search for the optimal   Borel parameters $T^2$ and continuum threshold parameters $s_0$  to satisfy   the  four criteria:\\
$\bf 1.$ Pole dominance at the hadron side;\\
$\bf 2.$ Convergence of the operator product expansion at the QCD side;\\
$\bf 3.$ Appearance of the Borel platforms;\\
$\bf 4.$ Satisfying the energy scale formula,\\
 via  try and error.

The pole contributions (PC) or ground state tetraquark contributions are defined by
\begin{eqnarray}
{\rm{PC}}&=&\frac{\int_{4m_{b}^{2}}^{s_{0}}ds\rho\left(s\right)\exp\left(-\frac{s}{T^{2}}\right)} {\int_{4m_{b}^{2}}^{\infty}ds\rho\left(s\right)\exp\left(-\frac{s}{T^{2}}\right)}\, ,
\end{eqnarray}
 this  definition is adopted in all the QCD sum rules. The $Z_b(10610)$, $Z_b(10650)$, $Z_c(3900)$ and $Z_c(4020)$ have analogous properties, the $Z_c(4430)$ can be tentatively assigned to be the first radial excited state of the $Z_c(3900)$ \cite{Wang-tetra-Zc4430,Azizi-Z4430,ChenHX-Z4430}. The relevant  mass gaps
 $M_{Z_c(4430)}-M_{Z_c(3900)}=591\,\rm{MeV}$, $M_{\psi^\prime}-M_{J/\psi}=589\,\rm{GeV}$, $M_{\Upsilon^\prime}-M_{\Upsilon}=563\,\rm{MeV}$ from the Particle Data Group \cite{PDG}. The energy gaps at the charmonium section have the relation $M_{Z^\prime_c}-M_{Z_c}=M_{\psi^\prime}-M_{J/\psi}$, we expect such a relation survives in
 the bottom section  $M_{Z^\prime_b}-M_{Z_b}=M_{\Upsilon^\prime}-M_{\Upsilon}=0.55\,\rm{GeV}$. In this article, we choose the continuum threshold parameters
 $\sqrt{s_0}=Z_b+0.55\pm0.10\,\rm{GeV}$ as a constraint.

 To judge the convergence of the operator product expansion, we calculate the contributions of the vacuum condensates $D(n)$
 with the formula,
\begin{eqnarray}
D(n)&=&\frac{\int_{4m_{b}^{2}}^{s_{0}}ds\rho_{n}(s)\exp\left(-\frac{s}{T^{2}}\right)}
{\int_{4m_{b}^{2}}^{s_{0}}ds\rho\left(s\right)\exp\left(-\frac{s}{T^{2}}\right)}\, ,
\end{eqnarray}
rather than with the formula,
\begin{eqnarray}
D(n)&=&\frac{\int_{4m_{b}^{2}}^{\infty}ds\rho_{n}(s)\exp\left(-\frac{s}{T^{2}}\right)}
{\int_{4m_{b}^{2}}^{\infty}ds\rho\left(s\right)\exp\left(-\frac{s}{T^{2}}\right)}\, . \label{Dninfty}
\end{eqnarray}
 The definition  in Eq.(\ref{Dninfty}) works only when all the contributions at the hadron side are included, such as the ground state, first radial excited state, second radial excited state, $\cdots$, continuum states, where the index $n$ denotes the dimension of the vacuum condensates.

 The energy scale formula $\mu=\sqrt{M^2_{X/Y/Z}-(2{\mathbb{M}}_Q)^2}$ for the QCD spectral densities can enhance the pole contributions remarkably and improve the convergence of the operator product expansion considerably.
In Ref.\cite{WangHuang-2014-NPA}, we take the energy scale formula $\mu=\sqrt{M^2_{X/Y/Z}-(2{\mathbb{M}}_b)^2}$ with the effective $b$-quark mass  ${\mathbb{M}}_b=5.13\,\rm{GeV}$ to determine the ideal energy scales of the QCD spectral densities. After its publication,  we re-checked the numerical calculations and found that there existed   a small error involving the mixed condensates. We corrected the small error and observed that the Borel windows were modified slightly and the predicted masses were  improved slightly,  but the conclusions survived, the updated effective $b$-quark mass was ${\mathbb{M}}_b=5.17\,\rm{GeV}$ \cite{WangZG-X5568}. In this article, we take the updated value ${\mathbb{M}}_b=5.17\,\rm{GeV}$. Moreover, we  recalculate the high dimensional vacuum condensates using the formula $t^a_{ij}t^a_{mn}=-\frac{1}{6}\delta_{ij}\delta_{mn}+\frac{1}{2}\delta_{jm}\delta_{in}$, and obtain slightly  different expressions  compared to
  the old calculations, where  $t^a=\frac{\lambda^a}{2}$, the $\lambda^a$ is the Gell-Mann matrix.

 We obtain the Borel parameters, continuum threshold parameters, energy scales of the QCD spectral densities,  pole contributions, and the contributions of the vacuum condensates of dimension $10$, which are shown explicitly in Table \ref{BorelP}.
From the Table,  we can see that the pole contributions are about $(44-66)\%$, the pole dominance condition  is well satisfied.

  In calculations, we observe that the contributions of the vacuum condensates of dimension $10$  $|D(10)|\leq 4 \%$ for the most QCD sum rules, the operator product expansion is well convergent. In the QCD sum rules for the tetraquark state $[ub]_{A}[\overline{db}]_{A}$ with $J^{PC}=0^{++}$, the $|D(10)|= (4\sim 11) \%$, which is somewhat large. If we choose Borel window $T^2=(6.7-7.7)\,\rm{GeV}^2$, we can obtain a lightly smaller pole contribution, ${\rm PC}=(40-63)\%$, and slightly smaller $D(10)$, $|D(10)|= (3\sim 8) \%$, which is acceptable. As the predicted mass changes slowly with variation of the Borel parameter, the Borel windows $T^2=(6.7-7.7)\,\rm{GeV}^2$ and $(6.4-7.4)\,\rm{GeV}^2$ lead to the same tetraquark mass.
  The two basic criteria of the QCD sum rules are satisfied.

We take into account all the uncertainties of the input parameters and obtain the masses and pole residues of the scalar, axialvector, tensor hidden-bottom  tetraquark states, which are shown explicitly in Table \ref{mass-Table} and in Figs.\ref{mass-1-fig}--\ref{residue-2-fig}. From  Tables \ref{BorelP}--\ref{mass-Table}, we can see that the energy scale formula $\mu=\sqrt{M^2_{X/Y/Z}-(2{\mathbb{M}}_b)^2}$ is well satisfied.
 In  Figs.\ref{mass-1-fig}--\ref{residue-2-fig}, we plot the masses and pole residues of the scalar, axialvector, tensor hidden-bottom tetraquark states with variations of the Borel parameters at much larger ranges than the Borel widows, the regions between the two
    perpendicular lines are the Borel windows,  where the Borel platforms appear.  Now the four criteria of the QCD sum rules are all satisfied.

 In Figs.\ref{mass-1-fig}--\ref{mass-2-fig}, we also present the experimental values of the masses of the $Z_b(10610)$ and $Z_b(10650)$ \cite{Belle1110,PDG}. From the figures, we can see that the  masses of all the
 $[ub]_S[\overline{db}]_{A}-[ub]_{A}[\overline{db}]_S$,
$[ub]_{A}[\overline{db}]_{A}$, $[ub]_{\widetilde{A}}[\overline{db}]_{A}-[ub]_{A}[\overline{db}]_{\widetilde{A}}$,
$[ub]_S[\overline{db}]_{\widetilde{A}}-[ub]_{\widetilde{A}}[\overline{db}]_S$ hidden-bottom tetraquark states with the $J^{PC}=1^{+-}$ are in excellent agreements  with the experimental values
$ M_{Z_b(10610)}=(10607.2\pm2.0)\,\rm{ MeV}$ and $M_{Z_b(10650)}=(10652.2\pm1.5)\,\rm{MeV}$ within uncertainties \cite{Belle1110,PDG}. We cannot assign the $Z_b(10610)$ and $Z_b(10650)$ unambiguously with the mass alone, we should study the partial decay widths exclusively to obtain a more robust assignment.

From Table \ref{mass-Table}, we can see that the scalar tetraquark states  $[ub]_{S}[\overline{db}]_{S}$, $[ub]_{A}[\overline{db}]_{A}$, $[ub]_{\tilde{A}}[\overline{db}]_{\tilde{A}}$, axialvector tetraquark states $[ub]_S[\overline{db}]_{A}-[ub]_{A}[\overline{db}]_S$,
$[ub]_{A}[\overline{db}]_{A}$,
 $[ub]_{\widetilde{A}}[\overline{db}]_{A}-[ub]_{A}[\overline{db}]_{\widetilde{A}}$,
$[ub]_S[\overline{db}]_{\widetilde{A}}-[ub]_{\widetilde{A}}[\overline{db}]_S$,
$[ub]_S[\overline{db}]_{A}+[ub]_{A}[\overline{db}]_S$,
$[ub]_{\widetilde{V}}[\overline{db}]_{V}-[ub]_{V}[\overline{db}]_{\widetilde{V}}$, and tensor tetraquark state
$[ub]_{A}[\overline{db}]_{A}$ have almost degenerated masses, i.e. about $10.6\,\rm{GeV}$. The calculations based on the QCD sum rules indicate that the scalar ($S$) and axialvector ($A$) bottom diquark states have degenerated masses \cite{WangDiquark}, furthermore,  the spin-spin interaction is proportional  to $\frac{\vec{S}_{i}\cdot \vec{S}_{j}}{m_i m_j}$, the mass of the $b$-quark is large, so it is reasonable  that the lowest scalar, axialvector, tensor  hidden-bottom tetraquark have almost degenerated masses.
The scalar diquark states $S$ and axialvector diquark states $A$, $\widetilde{A}$ are all good diquark states in building the lowest tetraquark states.

The hidden-bottom tetraquark masses obtained in the present work can be confronted to the experimental data  at the LHCb, Belle II,  CEPC, FCC, ILC
 in the future, and shed light on the nature of the exotic $X$, $Y$, $Z$ particles. We can take the pole residues as input parameters to study the two-body
 strong decays of those hidden-bottom tetraquark states
 \begin{eqnarray}
Z_b^+(1^{+-}) &\to&\pi^{+}\Upsilon({\rm 1,2,3S})\,  ,  \, \pi^{+} h_b({\rm 1,2P}) \, ,  \, \rho^{+} \eta_b ({\rm 1S})\, , \, (B\bar{B}^{*})^+\, ,\, (B^{*}\bar{B})^+\, ,\, (B^{*}\bar{B}^{*})^+\, , \nonumber\\
Z_b^+(0^{++}) &\to&\pi^{+}\eta_b({\rm 1,2S})\,  ,  \, \pi^{+} \chi_{b1}({\rm 1,2P}) \, ,  \, \rho^{+} \Upsilon({\rm 1S}) \, , \, (B\bar{B})^+\, ,\, (B^{*}\bar{B}^{*})^+\, , \nonumber\\
Z_b^+(1^{++}) &\to& \pi^{+} \chi_{b1}({\rm 1,2P}) \, ,  \, \rho^{+} \Upsilon({\rm 1S}) \, , \, (B\bar{B}^{*})^+\, ,\, (B^{*}\bar{B})^+\, ,\, (B^{*}\bar{B}^{*})^+\, , \nonumber\\
Z_b^+(2^{++}) &\to&\pi^{+}\eta_b({\rm 1,2S})\,  ,  \, \pi^{+} \chi_{b1}({\rm 1,2P}) \, ,  \, \rho^{+} \Upsilon({\rm 1S}) \, , \, (B\bar{B})^+\, ,\, (B^{*}\bar{B}^{*})^+\, ,
\end{eqnarray}
with the three-point QCD sum rules, and obtain the partial decay widths, and compare  them to the experimental data in the future to diagnose the nature of the $Z_b$ states, as different quark structures lead to different partial decay widths.
 In Ref.\cite{WangZhang-Solid},  we tentatively assign the $Z_c^\pm(3900)$ to be  the diquark-antidiquark  type axialvector  tetraquark  state,   study  the two-body strong decays $Z_c^+(3900)\to J/\psi\pi^+$, $\eta_c\rho^+$, $D^+ \bar{D}^{*0}$, $\bar{D}^0 D^{*+}$ with the QCD sum rules  based on solid quark-hadron quality, and obtain the total width of the $Z_c^\pm(3900)$, which  supports  assigning the $Z_c^\pm(3900)$ to be  the diquark-antidiquark  type axialvector  tetraquark  state.
 In Ref.\cite{Wang-Solid-Y4660}, we extend the method to study the two-body strong decays of the $Y(4660)$ as a diquark-antidiquark  type vector  tetraquark  state,
  and illustrate  how to study the relevant hadronic coupling constants   based on solid quark-hadron quality. The new method can be applied to study the two-body strong decays of the $Z_b$ tetraquark states directly.

The ratios of the partial widths of the decays $Z_c(3900/4020) \to \eta_c\rho\, , \,J/\psi \pi$ at $\sqrt{s}=4.23\,\rm{GeV}$  measured by the BESIII collaboration are
{\begin{eqnarray}
R&=& \frac{\Gamma(Z_c(3900)\to\eta_c\rho)}{\Gamma(Z_c(3900)\to J/\psi\pi)}=2.1\pm0.8, \nonumber\\
R^\prime &=& \frac{\Gamma(Z_c(4020)\to\eta_c\rho)}{\Gamma(Z_c(4020)\to h_c\pi)}<1.9\, ,
\end{eqnarray}
at the $90\%$ C.L.  \cite{CZYuan-Decay}.
More precise experimental data are still needed to examine the theoretical calculations. As far as the hidden-bottom axialvector tetraquark candidates $Z_b(10610)$ and $Z_b(10650)$  are concerned, the  partial widths have not been measured yet, the experimental data are scarce. More theoretical and experimental works are still needed to assign the $Z_b(10610)$ and $Z_b(10650)$ unambiguously.

\begin{table}
\begin{center}
\begin{tabular}{|c|c|c|c|c|c|c|c|c|}\hline\hline
 $Z_b$                                                & $J^{PC}$ & $T^2 (\rm{GeV}^2)$ & $\sqrt{s_0}(\rm GeV) $      &$\mu(\rm{GeV})$   &pole         &$|D(10)|$ \\ \hline

$[ub]_{S}[\overline{db}]_{S}$                         & $0^{++}$ & $7.0-8.0$          & $11.16\pm0.10$              &$2.40$            &$(44-66)\%$  &$\leq3\%$   \\

$[ub]_{A}[\overline{db}]_{A}$                         & $0^{++}$ & $6.4-7.4$          & $11.14\pm0.10$              &$2.30$            &$(44-68)\%$  &$\leq11\%$    \\

$[ub]_{\tilde{A}}[\overline{db}]_{\tilde{A}}$         & $0^{++}$ & $7.2-8.2$          & $11.17\pm0.10$              &$2.40$            &$(45-66)\%$  &$\leq4\%$    \\

$[ub]_{\tilde{V}}[\overline{db}]_{\tilde{V}}$         & $0^{++}$ & $11.4-12.8$        & $12.22\pm0.10$              &$5.40$            &$(44-61)\%$  &$\ll 1\%$   \\

$[ub]_S[\overline{db}]_{A}-[ub]_{A}[\overline{db}]_S$ & $1^{+-}$ & $7.0-8.0$          & $11.16\pm0.10$              &$2.40$            &$(44-66)\%$  &$<4\%$    \\

$[ub]_{A}[\overline{db}]_{A}$                         & $1^{+-}$ & $7.1-8.1$          & $11.17\pm0.10$              &$2.40$            &$(44-65)\%$  &$\leq4\%$  \\

$[ub]_{\widetilde{A}}[\overline{db}]_{A}-[ub]_{A}[\overline{db}]_{\widetilde{A}}$ & $1^{+-}$ & $6.9-7.9$     & $11.17\pm0.10$     &$2.40$      &$(44-66)\%$ &$\leq7\%$ \\

$[ub]_S[\overline{db}]_{\widetilde{A}}-[ub]_{\widetilde{A}}[\overline{db}]_S$     & $1^{+-}$ & $7.1-8.1$     & $11.17\pm0.10$     &$2.40$      &$(44-66)\%$ &$\leq4\%$ \\

$[ub]_S[\overline{db}]_{A}+[ub]_{A}[\overline{db}]_S$ & $1^{++}$ & $7.1-8.1$          & $11.18\pm0.10$              &$2.45$            &$(44-65)\%$  &$\leq3\%$   \\

$[ub]_{\widetilde{V}}[\overline{db}]_{V}-[ub]_{V}[\overline{db}]_{\widetilde{V}}$ & $1^{++}$ & $6.8-7.8$     & $11.19\pm0.10$     &$2.50$      &$(44-66)\%$ &$\leq4\%$ \\

$[ub]_{\widetilde{A}}[\overline{db}]_{A}+[ub]_{A}[\overline{db}]_{\widetilde{A}}$ & $1^{++}$ & $9.7-11.1$    & $11.99\pm0.10$     &$4.90$      &$(44-63)\%$ &$\ll 1\%$ \\

$[ub]_{A}[\overline{db}]_{A}$                         & $2^{++}$ & $7.2-8.2$          & $11.19\pm0.10$              &$2.50$            &$(44-65)\%$         &$<4\%$ \\ \hline\hline
\end{tabular}
\end{center}
\caption{ The Borel parameters, continuum threshold parameters, energy scales of the QCD spectral densities,  pole contributions, and the contributions of the vacuum condensates of dimension $10$  for the ground state hidden-bottom tetraquark states. }\label{BorelP}
\end{table}

\begin{table}
\begin{center}
\begin{tabular}{|c|c|c|c|c|c|c|c|c|}\hline\hline
 $Z_b$                                                                  & $J^{PC}$  & $M_Z (\rm{GeV})$   & $\lambda_Z (\rm{GeV}^5) $             \\ \hline

$[ub]_{S}[\overline{db}]_{S}$                                           & $0^{++}$  & $10.61\pm0.09$     & $(1.10\pm0.17)\times 10^{-1}$           \\

$[ub]_{A}[\overline{db}]_{A}$                                           & $0^{++}$  & $10.60\pm0.09$     & $(1.61\pm0.25)\times 10^{-1}$           \\

$[ub]_{\tilde{A}}[\overline{db}]_{\tilde{A}}$                           & $0^{++}$  & $10.61\pm0.09$     & $(1.81\pm0.27)\times 10^{-1}$           \\

$[ub]_{\tilde{V}}[\overline{db}]_{\tilde{V}}$                           & $0^{++}$  & $11.66\pm0.12$     & $3.03\pm0.31    $                      \\

$[ub]_S[\overline{db}]_{A}-[ub]_{A}[\overline{db}]_S$                   & $1^{+-}$  & $10.61\pm0.09$     & $(1.08\pm0.16)\times 10^{-1}$        \\

$[ub]_{A}[\overline{db}]_{A}$                                           & $1^{+-}$  & $10.62\pm0.09$     & $(1.07\pm0.16)\times 10^{-1}$           \\

$[ub]_{\widetilde{A}}[\overline{db}]_{A}-[ub]_{A}[\overline{db}]_{\widetilde{A}}$ & $1^{+-}$   & $10.62\pm0.09$    & $(2.12\pm0.31)\times 10^{-1}$    \\

$[ub]_S[\overline{db}]_{\widetilde{A}}-[ub]_{\widetilde{A}}[\overline{db}]_S$     & $1^{+-}$   & $10.62\pm0.09$    & $(1.08\pm0.16)\times 10^{-1}$    \\

$[ub]_S[\overline{db}]_{A}+[ub]_{A}[\overline{db}]_S$                   & $1^{++}$  & $10.63\pm0.09$     & $(1.17\pm0.17)\times 10^{-1}$        \\

$[ub]_{\widetilde{V}}[\overline{db}]_{V}-[ub]_{V}[\overline{db}]_{\widetilde{V}}$ & $1^{++}$   & $10.63\pm0.09$    & $(1.22\pm0.20)\times 10^{-1}$    \\

$[ub]_{\widetilde{A}}[\overline{db}]_{A}+[ub]_{A}[\overline{db}]_{\widetilde{A}}$ & $1^{++}$   & $11.45\pm0.14$    & $(8.52\pm1.02)\times 10^{-1}$    \\

$[ub]_{A}[\overline{db}]_{A}$                                           & $2^{++}$  & $10.65\pm0.09$     & $(1.72\pm0.24)\times 10^{-1}$      \\ \hline\hline
\end{tabular}
\end{center}
\caption{ The masses and pole residues of the ground state hidden-bottom tetraquark states. }\label{mass-Table}
\end{table}

\begin{figure}
\centering
\includegraphics[totalheight=6cm,width=7cm]{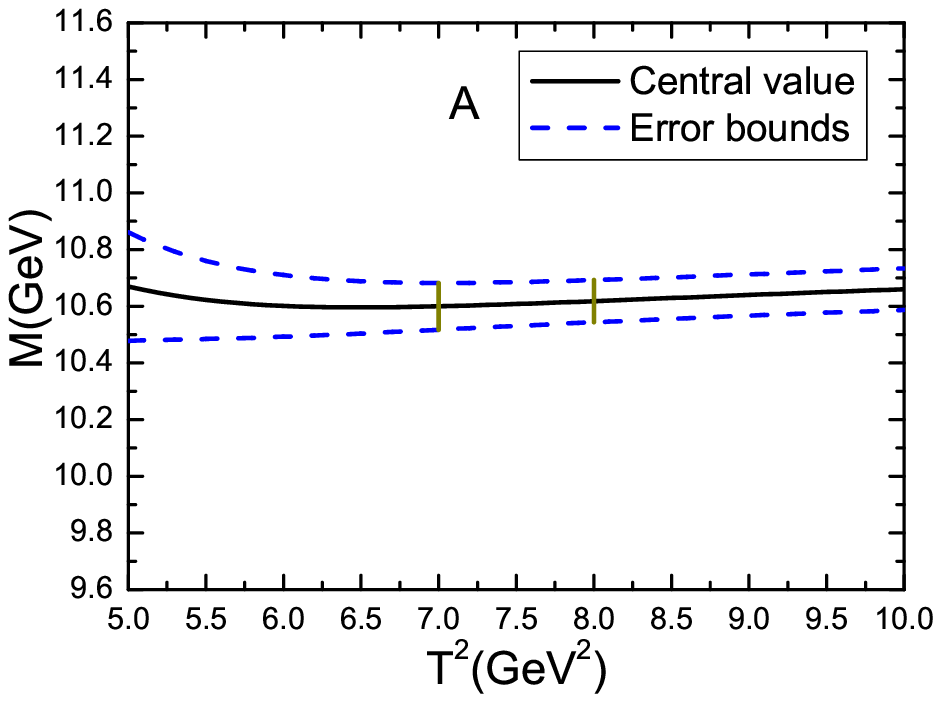}
\includegraphics[totalheight=6cm,width=7cm]{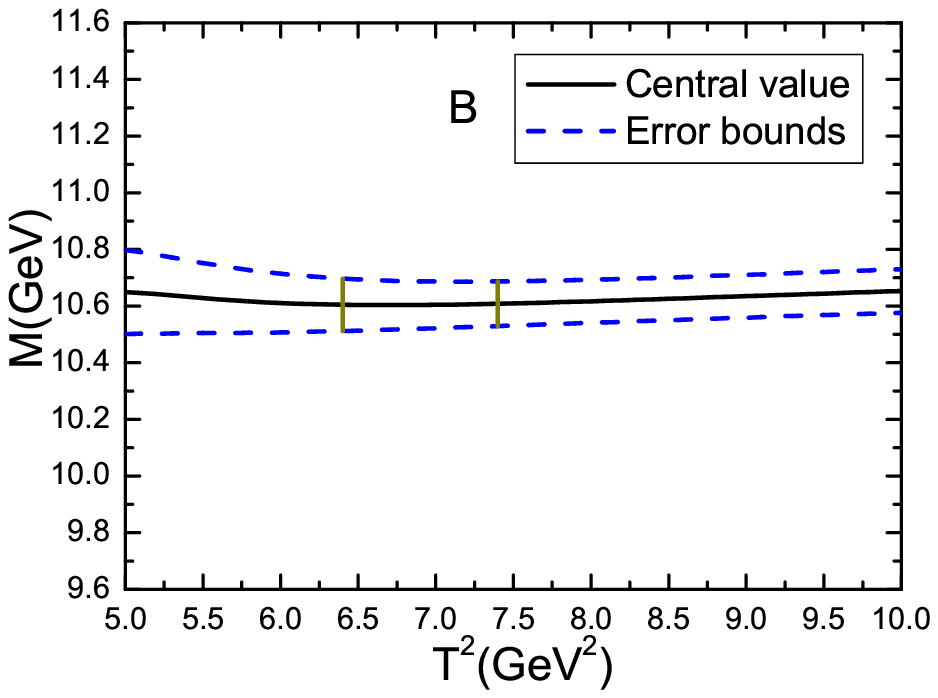}
\includegraphics[totalheight=6cm,width=7cm]{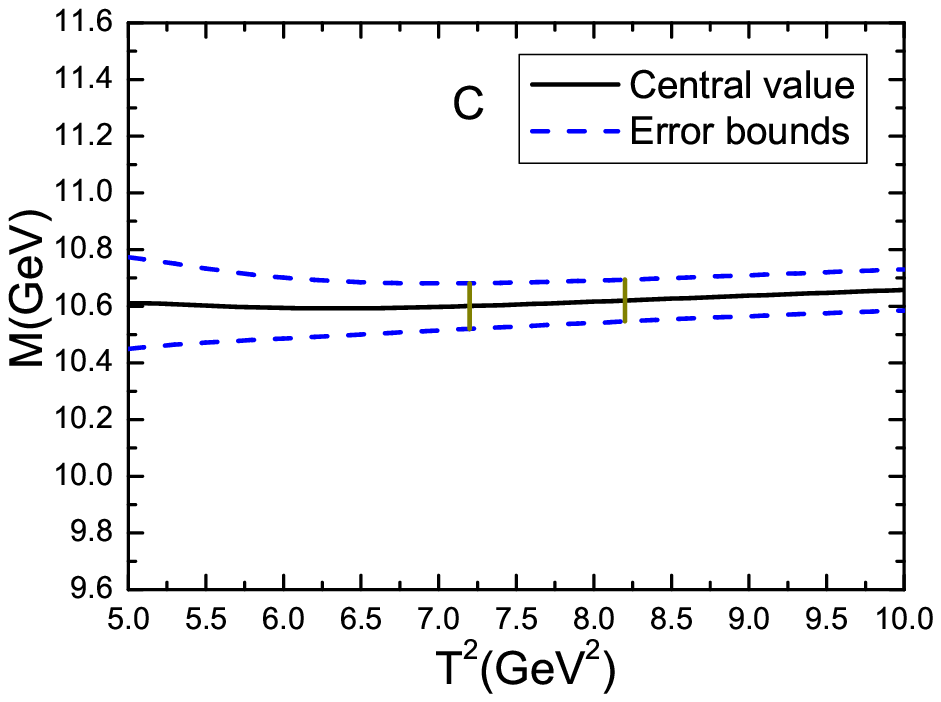}
\includegraphics[totalheight=6cm,width=7cm]{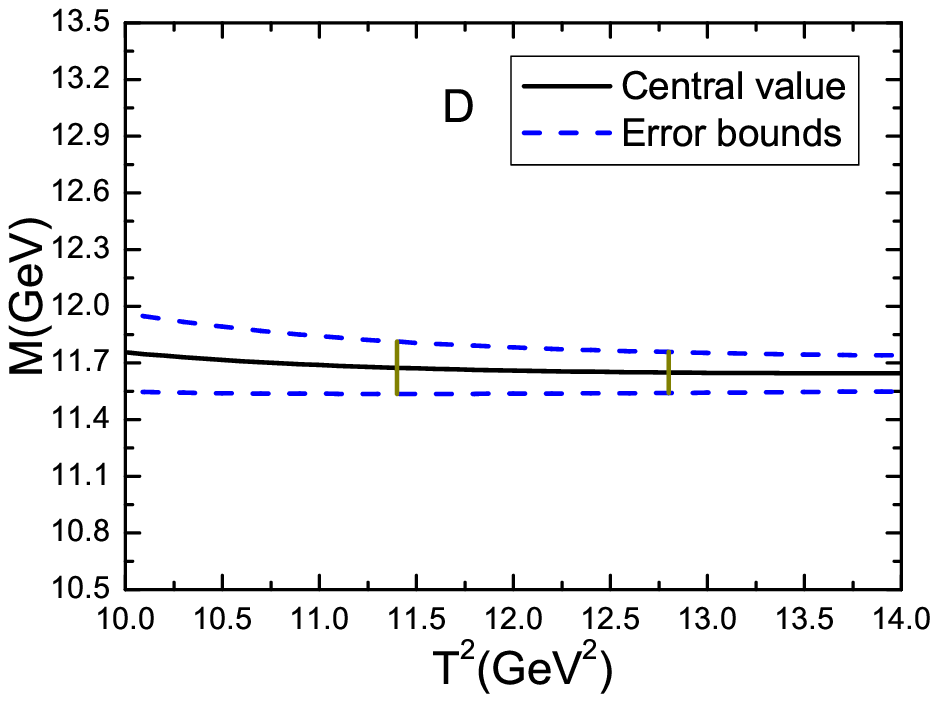}
\includegraphics[totalheight=6cm,width=7cm]{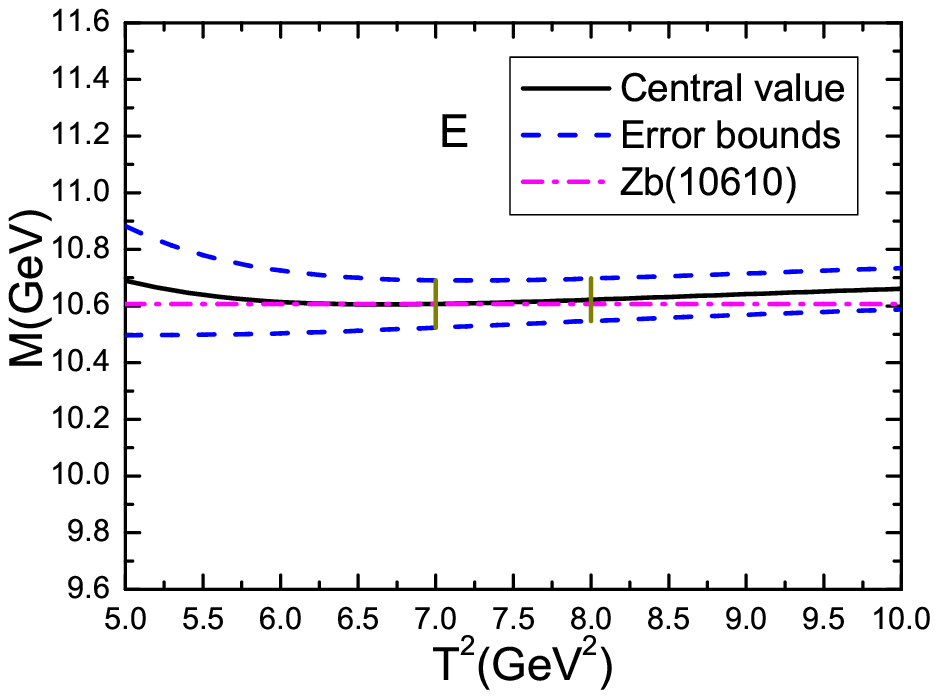}
\includegraphics[totalheight=6cm,width=7cm]{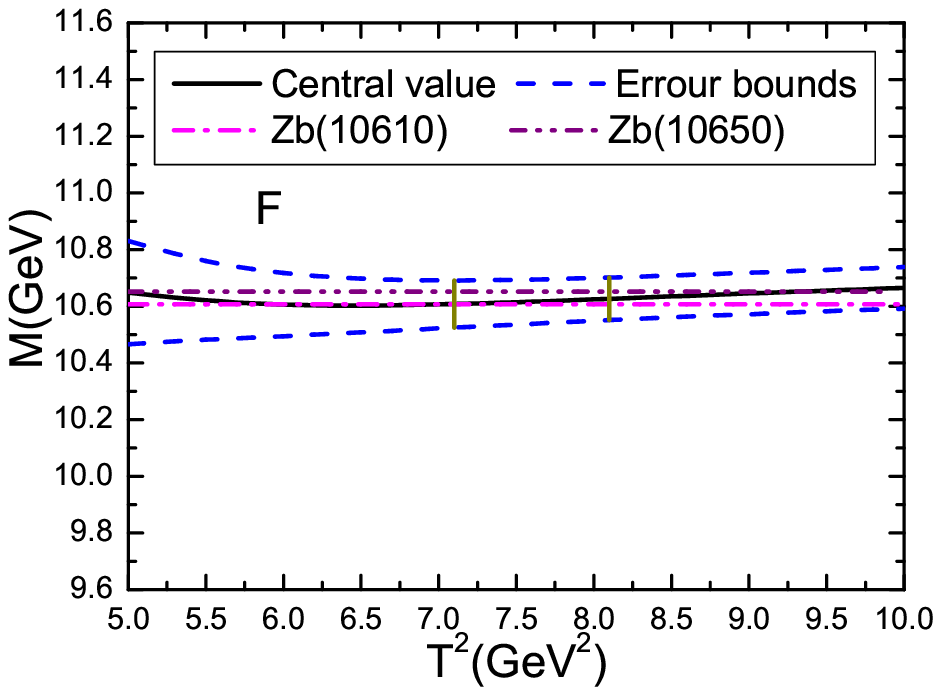}
  \caption{ The masses  with variations of the  Borel parameters $T^2$ for  the hidden-bottom tetraquark states, the $A$, $B$, $C$, $D$, $E$ and $F$  denote the
   tetraquark states   $[ub]_{S}[\overline{db}]_{S}$ ($0^{++}$),
$[ub]_{A}[\overline{db}]_{A}$ ($0^{++}$), $[ub]_{\tilde{A}}[\overline{db}]_{\tilde{A}}$ ($0^{++}$),
$[ub]_{\tilde{V}}[\overline{db}]_{\tilde{V}}$ ($0^{++}$),
$[ub]_S[\overline{db}]_{A}-[ub]_{A}[\overline{db}]_S$ ($1^{+-}$) and  $[ub]_{A}[\overline{db}]_{A}$ ($1^{+-}$), respectively, the regions between the two
     perpendicular lines are the Borel windows. }\label{mass-1-fig}
\end{figure}

\begin{figure}
\centering
\includegraphics[totalheight=6cm,width=7cm]{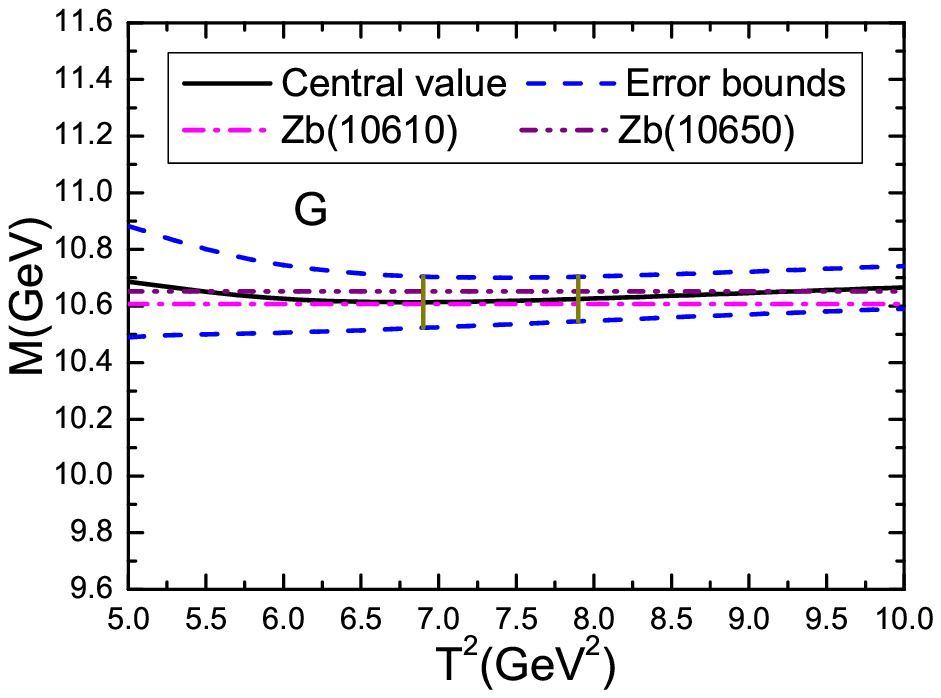}
\includegraphics[totalheight=6cm,width=7cm]{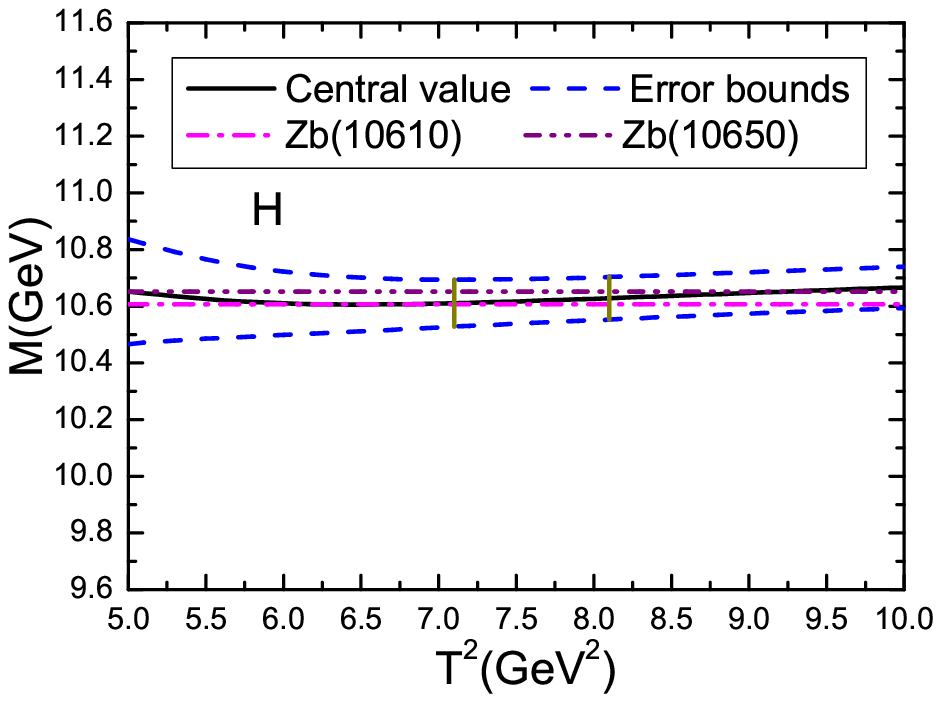}
\includegraphics[totalheight=6cm,width=7cm]{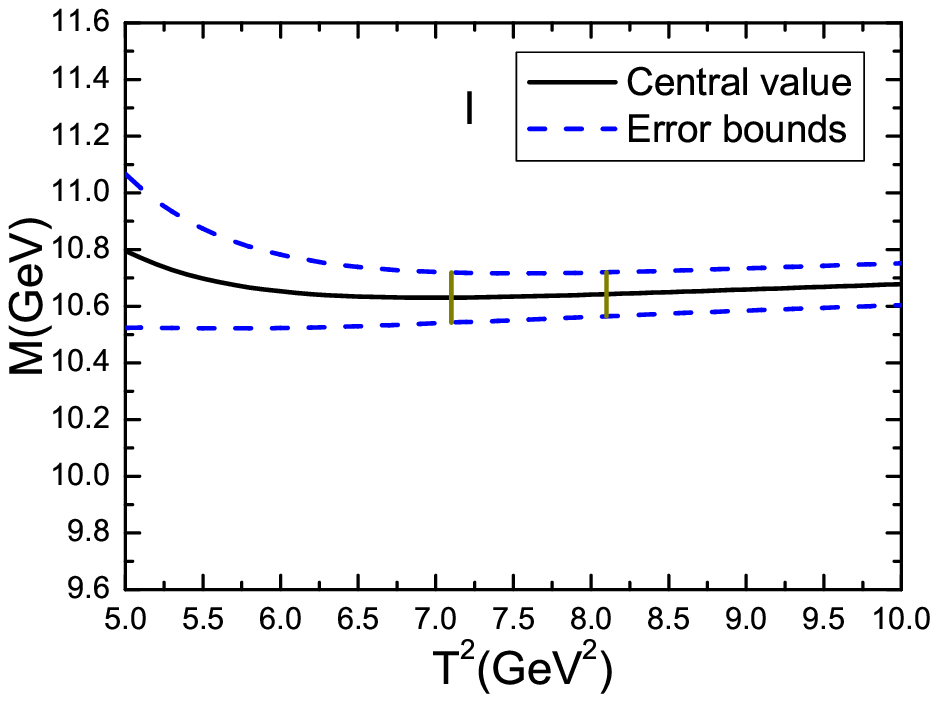}
\includegraphics[totalheight=6cm,width=7cm]{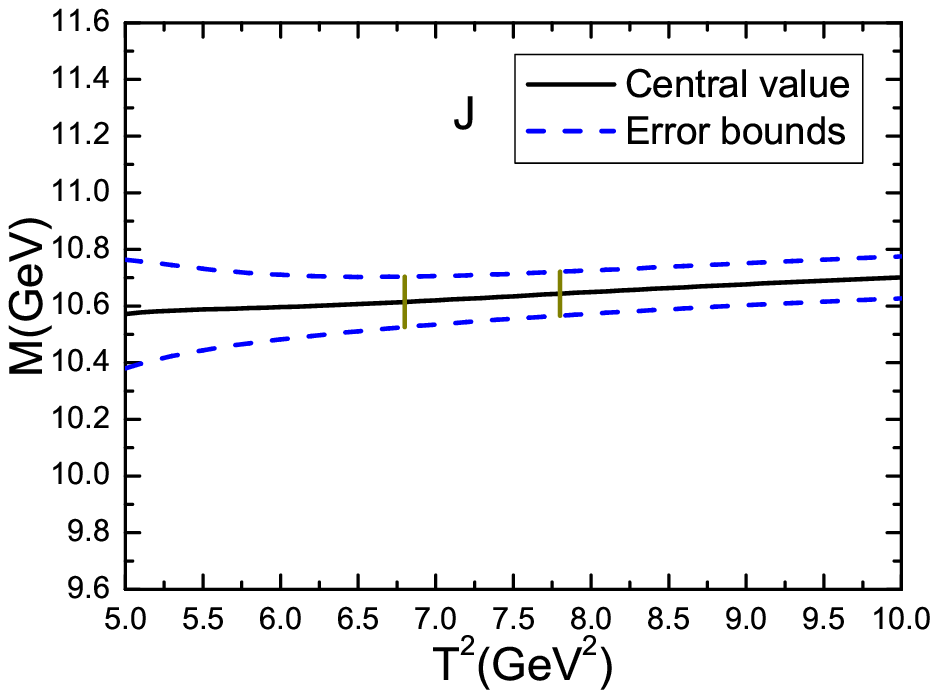}
\includegraphics[totalheight=6cm,width=7cm]{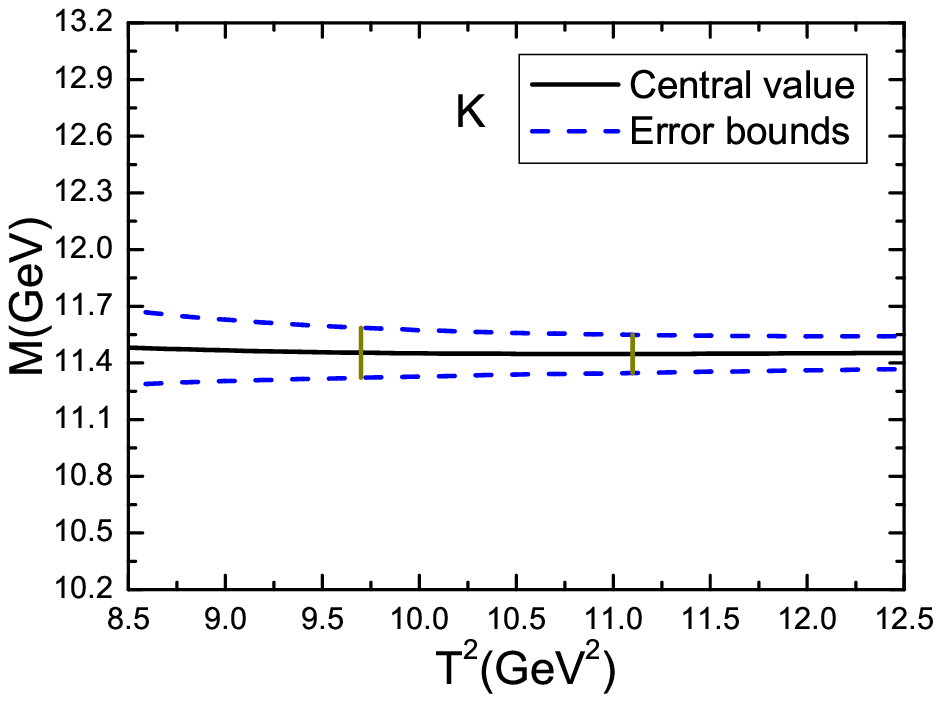}
\includegraphics[totalheight=6cm,width=7cm]{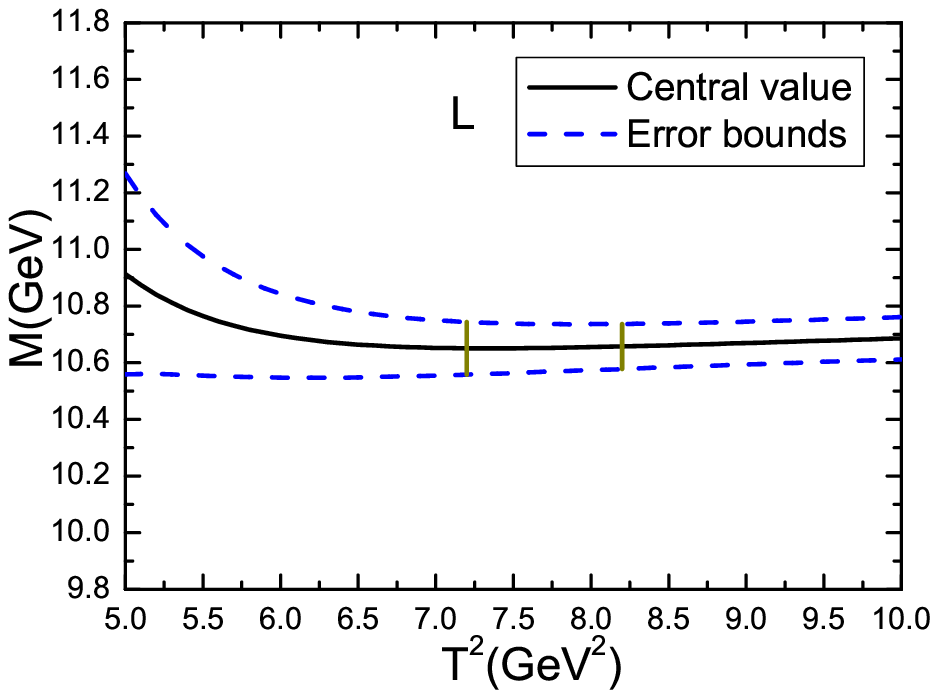}
  \caption{ The masses  with variations of the  Borel parameters $T^2$ for  the hidden-bottom tetraquark states, the $G$, $H$, $I$, $J$, $K$ and $L$  denote the
   tetraquark states  $[ub]_{\widetilde{A}}[\overline{db}]_{A}-[ub]_{A}[\overline{db}]_{\widetilde{A}}$ ($1^{+-}$),
$[ub]_S[\overline{db}]_{\widetilde{A}}-[ub]_{\widetilde{A}}[\overline{db}]_S$ ($1^{+-}$),
$[ub]_S[\overline{db}]_{A}+[ub]_{A}[\overline{db}]_S$ ($1^{++}$),
$[ub]_{\widetilde{V}}[\overline{db}]_{V}-[ub]_{V}[\overline{db}]_{\widetilde{V}}$ ($1^{++}$),
$[ub]_{\widetilde{A}}[\overline{db}]_{A}+[ub]_{A}[\overline{db}]_{\widetilde{A}}$ ($1^{++}$) and
$[ub]_{A}[\overline{db}]_{A}$ ($2^{++}$), respectively, the regions between the two
     perpendicular lines are the Borel windows. }\label{mass-2-fig}
\end{figure}

\begin{figure}
\centering
\includegraphics[totalheight=6cm,width=7cm]{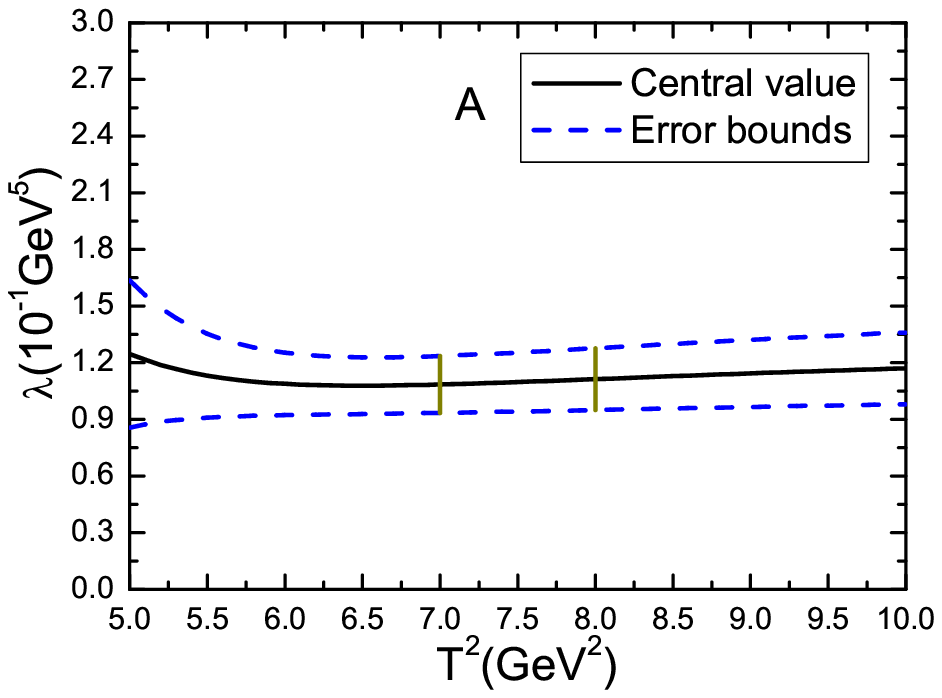}
\includegraphics[totalheight=6cm,width=7cm]{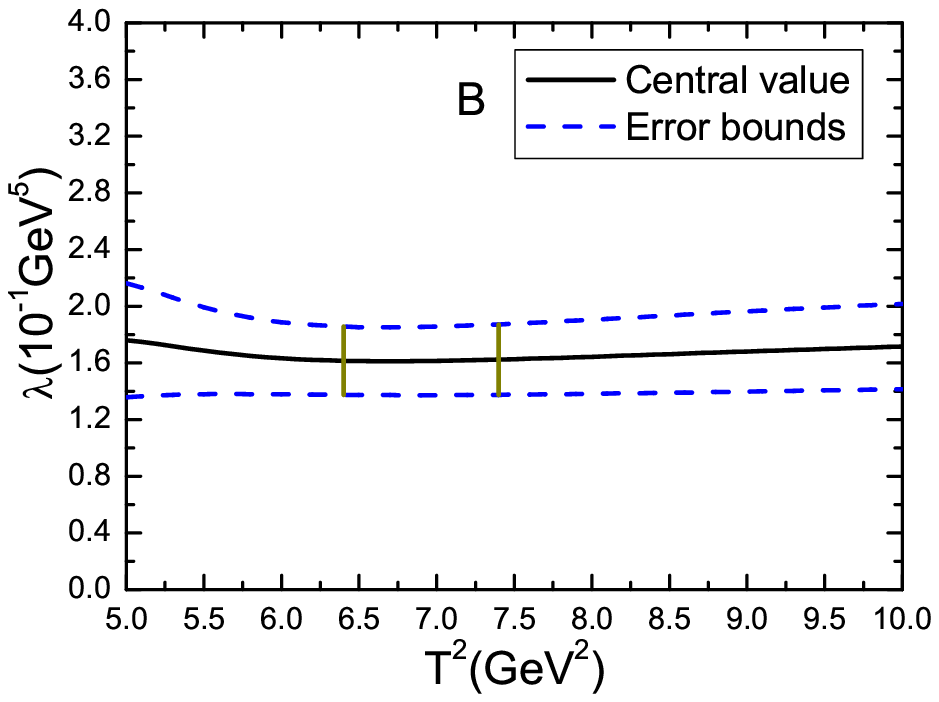}
\includegraphics[totalheight=6cm,width=7cm]{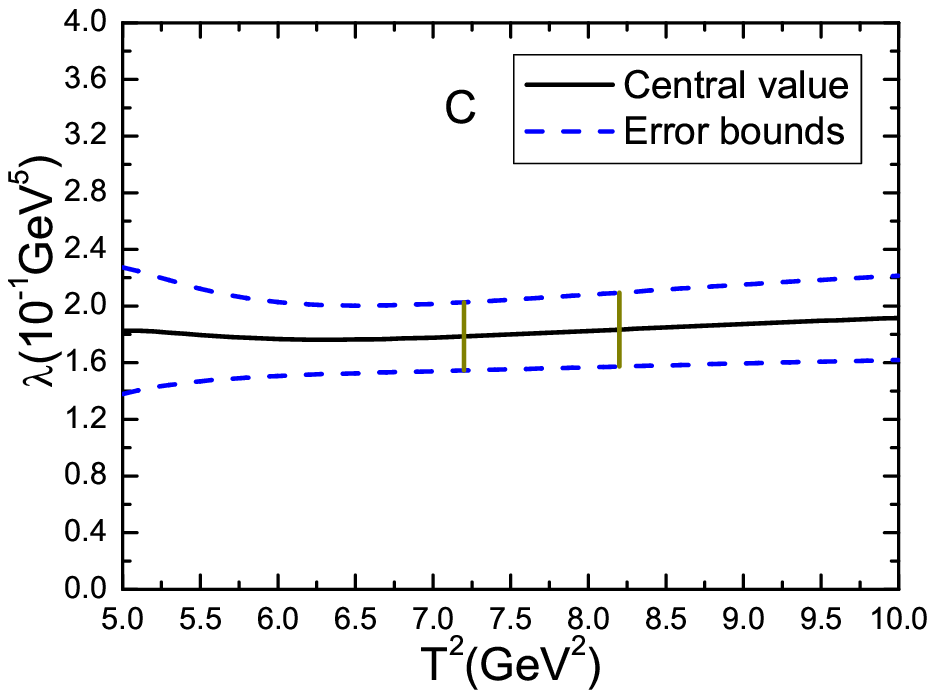}
\includegraphics[totalheight=6cm,width=7cm]{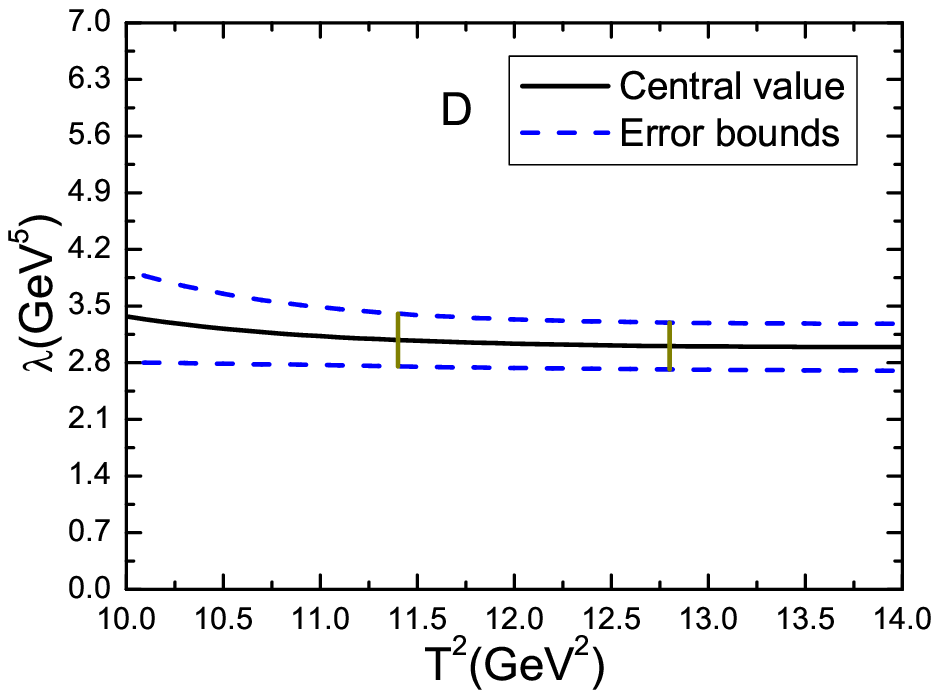}
\includegraphics[totalheight=6cm,width=7cm]{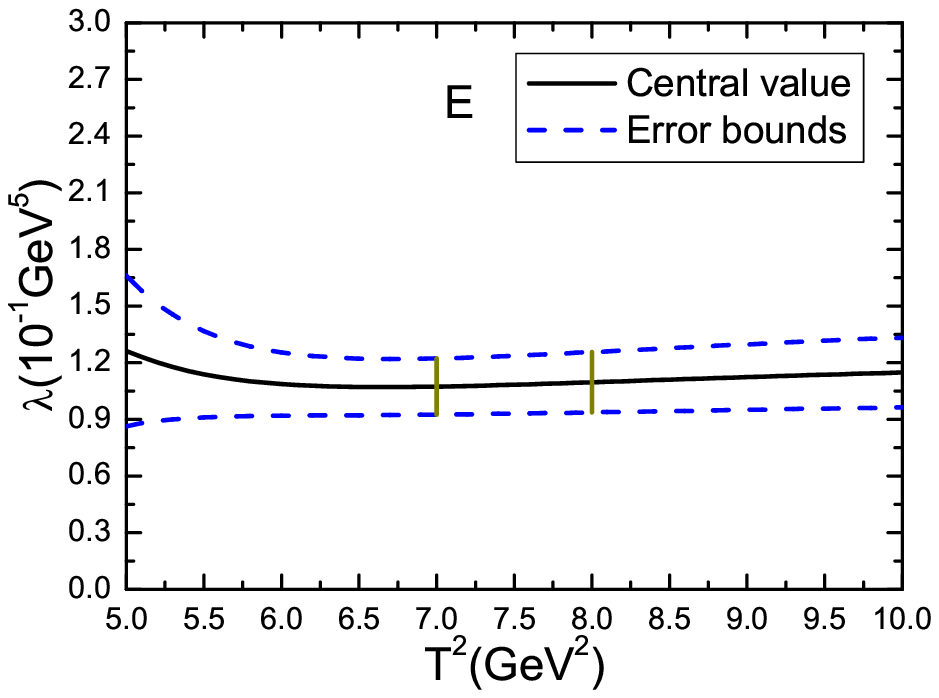}
\includegraphics[totalheight=6cm,width=7cm]{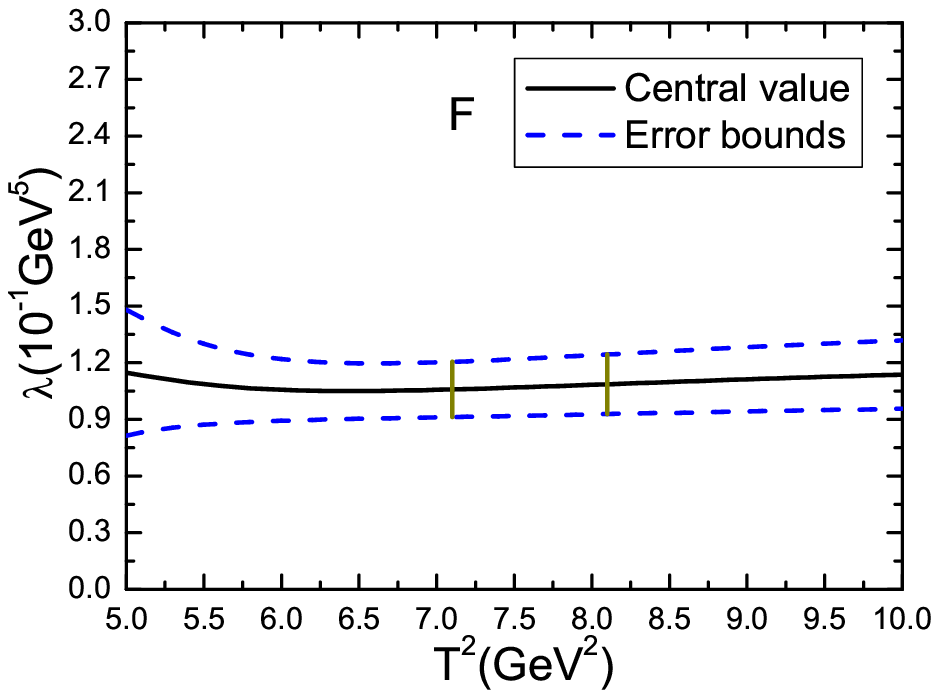}
  \caption{ The pole residues with variations of the  Borel parameters $T^2$ for  the hidden-bottom tetraquark states, the $A$, $B$, $C$, $D$, $E$ and $F$  denote the
   tetraquark states   $[ub]_{S}[\overline{db}]_{S}$ ($0^{++}$),
$[ub]_{A}[\overline{db}]_{A}$ ($0^{++}$), $[ub]_{\tilde{A}}[\overline{db}]_{\tilde{A}}$ ($0^{++}$),
$[ub]_{\tilde{V}}[\overline{db}]_{\tilde{V}}$ ($0^{++}$),
$[ub]_S[\overline{db}]_{A}-[ub]_{A}[\overline{db}]_S$ ($1^{+-}$) and  $[ub]_{A}[\overline{db}]_{A}$ ($1^{+-}$), respectively, the regions between the two
     perpendicular lines are the Borel windows.  }\label{residue-1-fig}
\end{figure}

\begin{figure}
\centering
\includegraphics[totalheight=6cm,width=7cm]{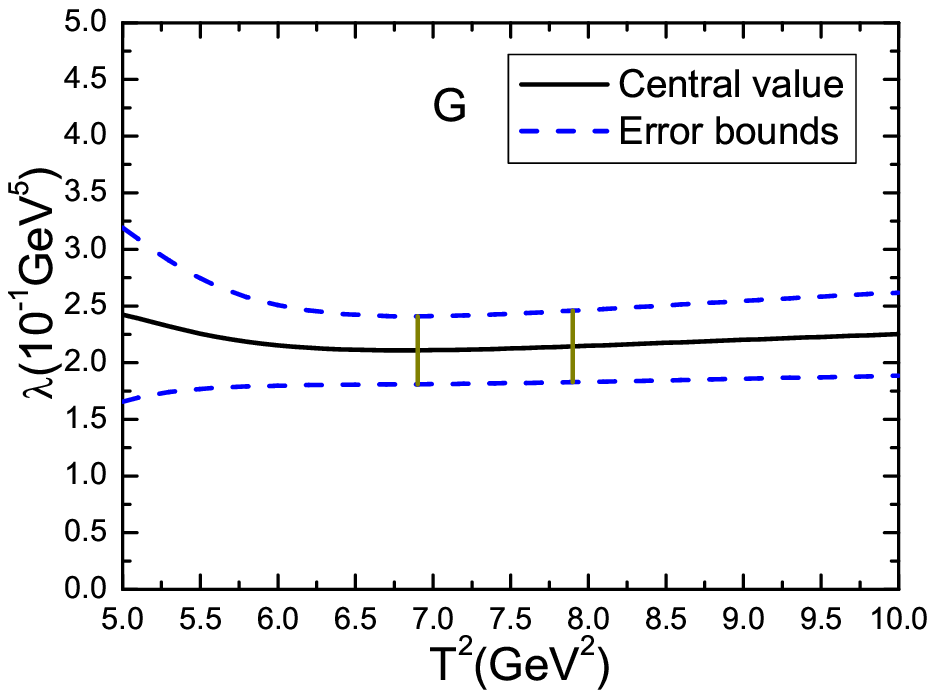}
\includegraphics[totalheight=6cm,width=7cm]{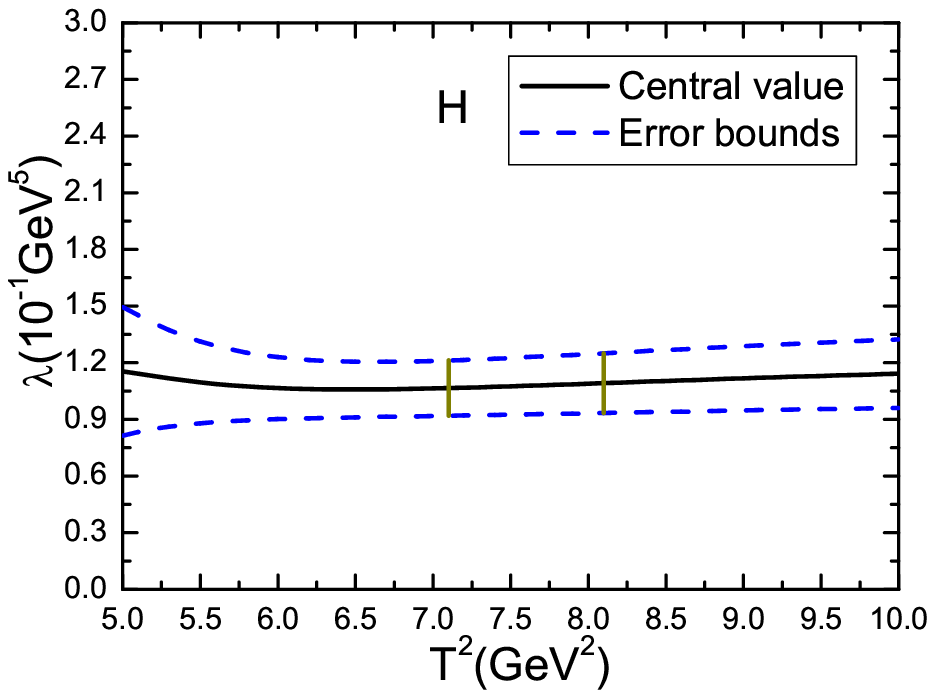}
\includegraphics[totalheight=6cm,width=7cm]{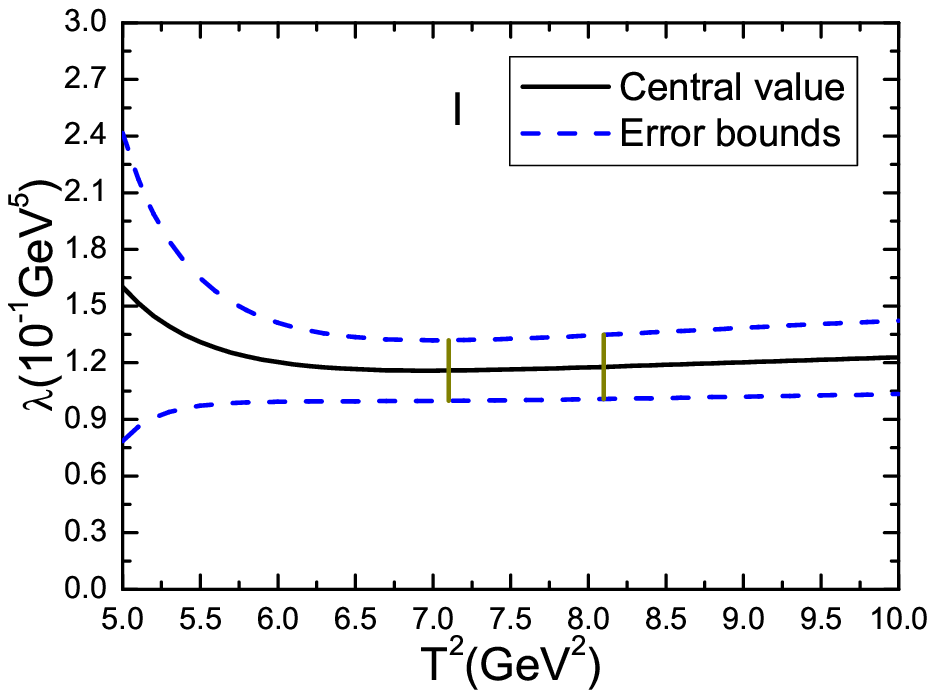}
\includegraphics[totalheight=6cm,width=7cm]{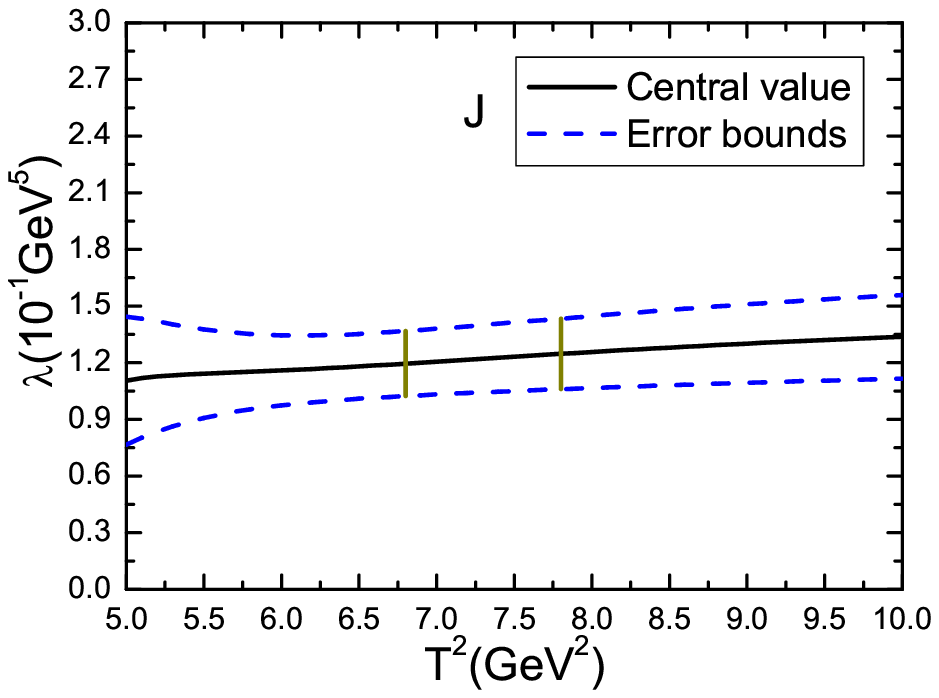}
\includegraphics[totalheight=6cm,width=7cm]{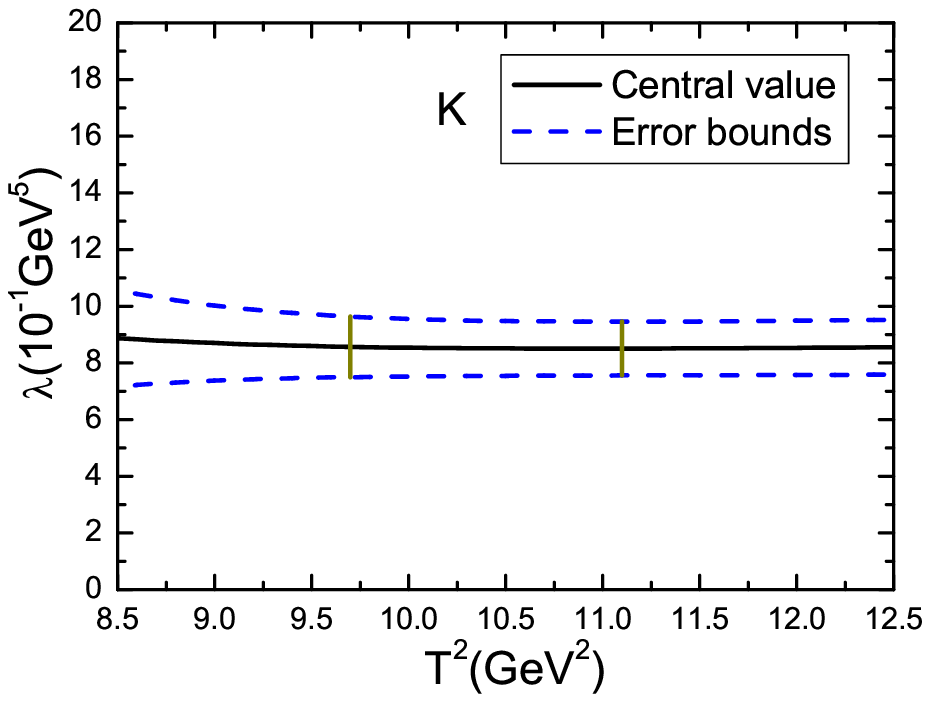}
\includegraphics[totalheight=6cm,width=7cm]{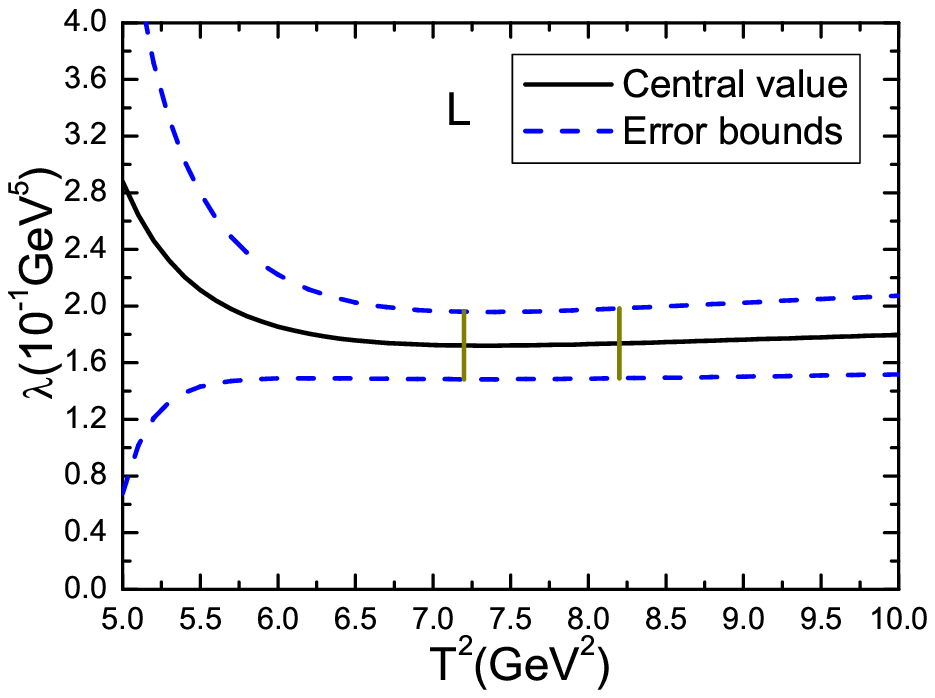}
  \caption{ The pole residues with variations of the  Borel parameters $T^2$ for  the hidden-bottom tetraquark states, the $G$, $H$, $I$, $J$, $K$ and $L$  denote the
   tetraquark states  $[ub]_{\widetilde{A}}[\overline{db}]_{A}-[ub]_{A}[\overline{db}]_{\widetilde{A}}$ ($1^{+-}$),
$[ub]_S[\overline{db}]_{\widetilde{A}}-[ub]_{\widetilde{A}}[\overline{db}]_S$ ($1^{+-}$),
$[ub]_S[\overline{db}]_{A}+[ub]_{A}[\overline{db}]_S$ ($1^{++}$),
$[ub]_{\widetilde{V}}[\overline{db}]_{V}-[ub]_{V}[\overline{db}]_{\widetilde{V}}$ ($1^{++}$),
$[ub]_{\widetilde{A}}[\overline{db}]_{A}+[ub]_{A}[\overline{db}]_{\widetilde{A}}$ ($1^{++}$) and
$[ub]_{A}[\overline{db}]_{A}$ ($2^{++}$), respectively, the regions between the two
     perpendicular lines are the Borel windows. }\label{residue-2-fig}
\end{figure}

\section{Conclusion}
In this article,  we take the scalar, axialvector and vector bottom (anti)diquark operators as the basic building blocks, and construct
  the scalar, axialvector and tensor hidden-bottom tetraquark currents to study the  mass spectrum of the ground state hidden-bottom tetraquark states  with
the QCD sum rules in a systematic way by carrying out the operator product expansion up to vacuum condensates of dimension $10$  consistently. In calculations, we use the energy scale formula $\mu=\sqrt{M^2_{X/Y/Z}-(2{\mathbb{M}}_b)^2}$ to determine the ideal energy scales of the QCD spectral densities and  choose the continuum threshold parameters  $\sqrt{s_0}=Z_b+0.55\pm0.10\,\rm{GeV}$ as a constraint to extract the masses and pole residues from the QCD sum rules. 
The predicted masses $10.61\pm0.09\,\rm{GeV}$ and $10.62\pm0.09\,\rm{GeV}$ for the $1^{+-}$ tetraquark states supports assigning the $Z_b(10610)$ and $Z_b(10650)$ to be the axialvector hidden-bottom tetraquark states, more theoretical and experimental works are still needed to assign the $Z_b(10610)$ and $Z_b(10650)$ unambiguously according to the partial decay widths.
The predicted tetraquark masses can be confronted to the experimental data in the future at the LHCb, Belle II,  CEPC, FCC, ILC. The pole residues can be taken as input parameters to study the two-body
 strong decays of those hidden-bottom tetraquark states with the three-point QCD sum rules.
Furthermore, we observe that the scalar diquark states $S$ and axialvector diquark states $A$, $\widetilde{A}$ are all good diquark states in building the lowest tetraquark states.

\section*{Acknowledgements}
This  work is supported by National Natural Science Foundation, Grant Number  11775079.

\end{document}